\documentclass[useAMS,usenatbib]{mn2e}
\pdfoutput=1
\usepackage{journals}
\usepackage[pdftex]{graphicx,color}
\usepackage[latin1]{inputenc}
\usepackage{graphics}
\usepackage{amsfonts}
\usepackage{amsmath}
\usepackage{multicol}
\usepackage{layout}
\usepackage{amssymb}
\usepackage{lipsum}
\usepackage[a4paper,colorlink   s=true,pdfstartview=FitV,
  linkcolor=red,citecolor=blue,urlcolor=magenta,draft]{hyperref}

 \title[Impact of baryonic physics on subhaloes]{The impact of baryonic physics
  on the subhalo mass function and implications for gravitational lensing}
\author[Despali    et     al.     2016]{\parbox{\textwidth}{
    Giulia Despali$^{1}$\thanks{E-mail:
      \href{mailto:gdespali@gmail.com}   {gdespali@gmail.com}}, Simona Vegetti$^{1}$\\}\\
  $^{1}$  Max  Planck   Institute  for
  Astrophysics, Karl-Schwarzschild-Strasse 1,  85740 Garching, Germany
  \\}

\begin{document}
\date{}
\maketitle
\label{firstpage}
\pagerange{\pageref{firstpage}--\pageref{lastpage}} \pubyear{2010}
\begin{abstract}
  We  investigate  the  impact  of baryonic  physics  on  the  subhalo
  population  by analyzing  the results  of two  recent hydrodynamical
  simulations  (EAGLE   and  Illustris),   which  have   very  similar
  configuration,  but  a  different  model of  baryonic  physics.   We
  concentrate  on   haloes  with   a  mass  between   $10^{12.5}$  and
  $10^{14}M_{\odot}h^{-1}$ and redshift between 0.2 and 0.5, comparing
  with  observational results  and  subhalo  detections in  early-type
  galaxy lenses.  We  compare the number and  the spatial distribution
  of subhaloes in  the fully hydro runs and in  their dark matter only
  counterparts,   focusing  on   the  differences   between  the   two
  simulations.   We find  that  the presence  of  baryons reduces  the
  number   of   subhaloes,   especially    at   the   low   mass   end
  ($\leq  10^{10}M_{\odot}h^{-1}$), by  different amounts  depending on
  the model.  The variations in the subhalo mass function are strongly
  dependent on  those in the halo  mass function, which is  shifted by
  the effect of  stellar and AGN feedback. 
  Finally, we search  for analogues of the observed  lenses (SLACS) in
  the simulations, selecting them  in velocity dispersion and
  dynamical  properties.  We  use  the selected  galaxies to  quantify
  detection  expectations  based on  the  subhalo  populations in  the
  different simulations, calculating the detection probability and the
  predicted values for the  projected dark matter fraction in
  subhaloes   $f_{DM}$   and   the   slope  of   the   mass   function
  $\alpha$. We  compare these  values with those  derived from
    subhalo  detections   in  observations   and  conclude   that  the
    dark-matter-only  and hydro  EAGLE runs  are both  compatible with
    observational results,  while results from the  hydro Illustris run
    do not lie within the errors.

\end{abstract}

\section{introduction}

Numerical  simulations of  galaxy formation  are now  able to  produce
realistic galaxy populations, that  reproduce observed relations quite
well   \citep{vogel14,schaye15}.   Simulations   are  fundamental   to
understand  the  physical properties  that  shape  galaxies and  their
evolution;  while   the  cold-dark-matter-only  simulations   and  the
treatment  of  the   dark  matter  component,  in   general,  is  well
established, there  is still  not a  general consensus  on the
  details of  the baryonic physics implementation  and differences on
this side can lead to quite different predictions in terms of feedback
processes and the details of galaxy formation.  Moreover, non standard
descriptions of  the dark matter  component, such as warm  dark matter
(WDM)    models,     may    also    have    an     important    impact
\citep{lovell12,lovell14,li16}.

In this  work we analyse the  main runs of the  EAGLE \citep{schaye15}
and Illustris \citep{vogel14} projects, to investigate the effects
of different  baryonic models  on the substructure  population. Strong
gravitational lensing  allows us  to  detect directly the  presence of
substructures either via their effect on the relative flux of multiply
imaged quasars  \citep{dalal02,nierenberg14} or via their  effect on
the   surface   brightness  of   Einstein   rings   and  lensed   arcs
\citep{vegetti09,vegetti10,vegetti12,vegetti14,hezaveh16}.   Numerical
simulations   can  then   be  used   to  make   predictions  for   the
interpretation of  observational results,  and possibly rule  out dark
matter and galaxy formation models.

Previous works concerning subhaloes mainly study cold dark matter-only
simulations,   whose    results   are   nowadays    well   established
\citep{giocoli08a,springel10}.   Also, studies  investigating
in  detail subhaloes  and  their  evolution/distribution in  different
environments and dark matter/hydrodynamical models usually concentrate
on  Milky Way  haloes,  aiming  to address  the  well known  ``missing
satellites''     and     ``too      big     to     fail''     problems
\citep{dicintio11,dicintio13,garrison14,wetzel16}.   The  aim of  this
work is to  investigate the effect of baryonic physics  on the subhalo
population  concentrating on  haloes  between $10^{12.5}$  and
  $10^{14}M_{\odot}  h^{-1}$: this  corresponds  to the  halo mass  of
  massive  early-type  galaxies  (ETGs), which  act  as  gravitational
  lenses    in    the    recent   cases    of    subhalo    detections
  \citep{vegetti10,vegetti12}. In particular, we  want to focus on the
  differences  that can  arise  from different  models, stressing  the
  importance of an accurate implementation of baryonic physics.

The paper  is structured as  follows: we describe the  simulations and
our halo  selection in Section  \ref{sec_sim}.  First, we  analyse and
model  the  subhalo  mass   function  in  the  different  simulations,
concentrating on the difference in the number of subhaloes between the
dark  matter  only and  the  full  hydro runs  (Section  \ref{sec_mf},
\ref{sec_count}).   We  proceed  by  comparing  the  predictions  from
simulation   with  the   observational  results   and  what   are  the
probabilities of  detecting a substructure given  the predictions from
different  kind of  simulations:  in Section  \ref{sec_obs} we  select
analogues of observed systems and  in Section \ref{sec_det} we compare
the  detection  probability  inferred  from simulations  with  a  real
detection  in  an observational  sample  (SLACS  lenses).  In  Section
\ref{conclusion}  we   summarise  our  results.    Finally  in
  Appendix \ref{Struct}, we point out  the differences in the baryonic
  composition  of  haloes and  subhaloes  between  the EAGLE  and  the
  Illustris  simulations.  This difference  is  expected  to have  an
important impact on the gravitational  lensing effect of the subhaloes
and  their  detectability.  We  will investigate  this  further  in  a
follow-up paper.

\section{Simulations} \label {sec_sim}

We  choose  to analyse  the  main  runs  of  the EAGLE  and  Illustris
simulations  for many  reasons.  The  simulations have  comparable box
sizes, resolutions and starting  redshifts; moreover a dark-matter-only
counterpart, created with the same  initial conditions, exists in both
cases and  thus constitute an  ideal sample for comparison.   The main
papers from  the Illustris \citep{vogel14} and  EAGLE \citep{schaye15}
collaborations illustrate in detail the differences between the models
of baryonic  physics, in addition  to the  differences in the  codes -
AREPO   \citep{springel10}  and   a   modified   version  of   GADGET3
\citep{springel08a}, respectively  - used to run  the simulations.  It
has  been   shown  that,  when  looking   at  the  structural
properties of haloes in detail,  small but significant differences may
arise, caused by the simulation code \citep{heitmann08} or even by the
halo  \citep{knebe11,knebe13}  and subhalo  finders  \citep{onions12}.
These  variations  become  particularly  important when  we  focus  on
individual  structures or  small-scale  detail.   Similar versions  of
SUBFIND  \citep{springel01b} have  been  used in  both simulations  to
identify structures, eliminating part  of these potential differences.
Nevertheless, since  the baryonic component is  treated differently in
the  two  codes,  we  still   expect  some  effect  on  the  (sub)halo
identification.  We make use of the existing SUBFIND catalogues of the
two  simulations  and  we  concentrate  in the  mass  bin  of  massive
early-type  galaxies  (ETGs).    SUBFIND  starts  the  subhalo
  identification  from overdensity  peaks  within the  main halo (see
  \citealt{muldrew10} for  more details on  the algorithm), making  it a
  good candidate for a comparison  with the potential corrections used
  to    find    subhaloes     in    gravitational    lens systems
  \citep{vegetti10,vegetti09}.

Due  to the  resolution limits,  the  smallest subhaloes  have a  mass
$\simeq  10^{8}M_{\odot}h^{-1}$  (with  at least  10  particles).   We
include these subhaloes for statistical  purposes, but we caution that
reliable measurements require  a minimum of 100  particles per subhalo
\citep{onions12}.   The cases  where particle  numbers drop  below 100
particles will be  marked by a gray region when  necessary.  Given the
good agreement between the  two dark-matter-only (hereafter DMO) runs,
in this work we show results only  from EAGLE for the DMO case. We use
different halo mass definitions  throughout this work: $(i)$ $M_{200}$
is defined as the mass of a  sphere centered on the halo and enclosing
200     times     the     critical    density     $\rho_{c}$     (2.77
$\times 10^{11}M_{\odot}h^{-1}Mpc^{-3}$)  and we chose it  as the main
halo  mass   definition  since  it   is  more  easily   comparable  to
observational results  - thus  we will  refer to  this where  no other
definition is specified ; $(ii)$ $M_{fof}$ is the mass of the group in
the  halo catalogue  identified by  the  FOF algorithm,  which has  no
pre-defined shape  and is usually  larger than $M_{200}$:  this method
uses a  linking length  (conventionally set  to $b=0.2$)  to establish
which  particles  belong to  the  halo;  $(iii)$ $M_{vir}$  is
  defined  as  the  mass  within   the  sphere  enclosing  the  virial
  overdensity, which  is calculated from the  spherical collapse model
  \citep{bryan98}    and   depends    on   the    cosmological   model
  \citep{sheth99b,despali16}.  The  last two definitions will  be used
  respectively in Figure  \ref{massfunc} - as a  general definition to
  include  all haloes  and subhaloes  in the  catalogues -  and Figure
  \ref{slacs2} -  for a comparison  with the virial  masses calculated
  from the observational data.

We now list the main features of the two simulations.

\subsection{EAGLE}
In this work we analyse the main run of the EAGLE project, created as
part of a  Virgo Consortium project called the  Evolution and Assembly
of Galaxies and  their Environment \citep{schaye15,crain15,mcalpine16}
using a modified version of GADGET-3.  \citep{springel08b}.  The EAGLE
project consists  of simulations of $\Lambda$CDM  cosmological volumes
with  sufficient  size  and  resolution to  model  the  formation  and
evolution of  galaxies with a wide  range of masses, and  also includes a
counterpart set of dark matter only simulations of these volumes.  The
galaxy formation simulations include the correct proportion of baryons
and model gas hydrodynamics and radiative cooling and state-of-the-art
subgrid  models  are  used  to  follow  star  formation  and  feedback
processes  by both  stars and  AGN.  The  parameters of  the subgrid
model  have been  tuned to  match some  observational results,  as the
$z\simeq 0$  galaxy stellar  mass function  and the  observed relation
between stellar  and black hole  mass \citep{schaye15}.  The  main run
and its dark matter only counterpart follow $1504^{3}$ dark matter and
(in the first case) $1504^{3}$ gas particles in a box size of 100 Mpc,
from  redshift   $z=127$  to  the  present   time.   The  cosmological
parameters  were  set   to  the  best  fit  values   provided  by  the
\citet{planck1_14}         and        are:         $\Omega_{m}=0.307$,
$\Omega_{\Lambda}=0.693$,    $\Omega_{b}=0.04825$,    $h=0.677$    and
$\sigma_{8}=0.8288$.  With  this model, the dark  matter particle mass
is 1.15 $\times10^{7}M_{\odot}$  in the DMO run and 9.70
$\times10^{6}M_{\odot}$  in  the  full  one,  while  the  initial  gas
particle mass is 1.81 $\times10^{6}M_{\odot}$.

The galaxy formation model employs only one type of stellar feedback,
which captures  the collective  effects of  processes such  as stellar
winds, radiation pressure on dust  grains, and supernovae, and also only one
type of AGN  feedback (as opposed to e.g. both  a ``radio'' and ``quasar''
mode). Thus, as detailed below, it differs from the feedback
implementation of the Illustris.

\subsection{Illustris}

The Illustris  Project is  a series  of hydrodynamical  simulations of
cosmological volumes that follow the  evolution of dark matter, cosmic
gas, stars, and super massive black  holes from a starting redshift of
z  = 127  to the  present time.  In this  work, we  used the  main run
Illustris-1 (and the dark matter only run Illustris-1-Dark), which has
a box size  of 106.5 Mpc and follows $1820^{3}$  dark matter particles
and $1820^{3}$ (initial) gas cells. The simulations were run using the
recent  moving-mesh   AREPO  code  \citep{springel10}.    The  adopted
cosmological model has $\Omega_{m}=0.2726$, $\Omega_{\Lambda}=0.7274$,
$\Omega_{b}=0.0456$, $h=0.704$ and $\sigma_{8}=0.809$, consistent with
the WMAP-9 measurements \citep{wmap9}.  In  this case, the dark matter
particle mass  is 7.5 $\times10^{6}M_{\odot}$ in  the dark matter-only
run and 6.3 $\times10^{6}M_{\odot}$ in the full one, while the initial
gas particle  mass is 1.3 $\times10^{6}M_{\odot}$.  The  galaxy formation
model  includes gas  cooling (primordial  and metal  line cooling),  a
sub-resolution   ISM  model,   stochastic   star  formation,   stellar
evolution,  gas   recycling,  chemical  enrichment,   kinetic  stellar
feedback  driven  by  SNe,  procedures  for  supermassive  black  hole
seeding, super massive  black hole accretion and  merging, and related
AGN feedback  (radio-mode, quasar-mode,  and radiative) and  some free
parameters  constrained  based on  the  star  formation efficiency  in
smaller scale  simulations \citep{vogel14}. This  implementation leads
to a generally stronger AGN feedback with respect to the one in EAGLE.
We  made use  of the  data products  and the  scripts provided  in the
Illustris data release \citep{nelson15}.

 \section{The subhalo mass function} \label{sec_mf}

 The subhalo  (and also the  halo) mass function has  been extensively
 studied in  previous works \citep{giocoli08a,springel10},  which made
 use of dark matter only simulations.  It is well established that the
 presence of  baryons modifies  the halo  structure \citep{schaller15}
 and  may also  influence the  subhalo population,  in terms  of their
 number, spatial  and mass distribution.  A  direct comparison between
 DMO and hydrodynamical  simulations with this purpose can  be done by
 means  of  zoom-in  resimulations  \citep{zhu16,fiacconi16}  or  full
 cosmological  runs  (e.g.  \citealt{sawala13}) .   The  first  approach
 allows to reach very high  resolutions and thus small subhalo masses,
 but the  results are  intrinsically limited in  the number  of parent
 haloes and  thus do  not give statistical  predictions on  the subhalo
 mass function, while in the second case there are stronger limitation
 in  resolution;  on the  other  hand,  using full  cosmological  runs
 guarantees to take into account large  scale effect such as the total
 abundance  of haloes.  We use the  main EAGLE  and
   Illustris   runs  where the   lowest   subhalo    mass   is around
   $10^{8}M_{\odot}h^{-1}$; given  their very  similar
     overall  configuration,  they  allow  us  to  directly  test  how
     different baryonic physics  implementations influence the subhalo
     mass function.

 Generally, the  action of baryons  on the overall DM  distribution in
 haloes  and subhaloes  is threefold:  $(i)$ reionization  affects the
 formation  and  evolution  of  low-mass haloes,  making  them  almost
 completely     dark     (with      no     star     formation)     for
 $M<10^{9}M_{\odot}h^{-1}$; $(ii)$  the DM concentration in  the inner
 region is  increased due  to gas  cooling and  adiabatic contraction,
 both  in the  haloes and  in the  most massive  subhaloes which  host
 stars; $(iii)$  stellar and  AGN feedback  cause differences  in halo
 mass,     both     at    the     low     and     high    mass     end
 \citep{cui12,sawala13,vogel14,velliscig14,schaller15,sawala15},
 generally  making  haloes ``lighter''  in  the  hydro runs  and  thus
 shifting the  halo mass function; finally,  different feedback models
 may  affect how  matter  is stripped  from  subhaloes.  For  example,
 stronger  tidal  forces  in  the  hydro runs  may  remove  mass  more
 efficiently  from   subhaloes  during  their   infall  \citep{zhu16}.
 Earlier works \citep{sawala13,schaller15,vogel14} showed that
   the  abundances  of   both  haloes  and  subhaloes   are  reduced  in
   hydrodynamic  simulations compared  to the  DMO counterparts,  as a
   consequence of  a reduction in  mass.  Nevertheless, the  amount of
   this reduction  depends on the  baryonic physics model and  thus in
   the following  sections we will  explore in detail  the differences
   between the EAGLE and the Illustris runs.

\subsection{Halo vs. subhalo mass function}
\begin{figure*}
\includegraphics[width=0.7\hsize]{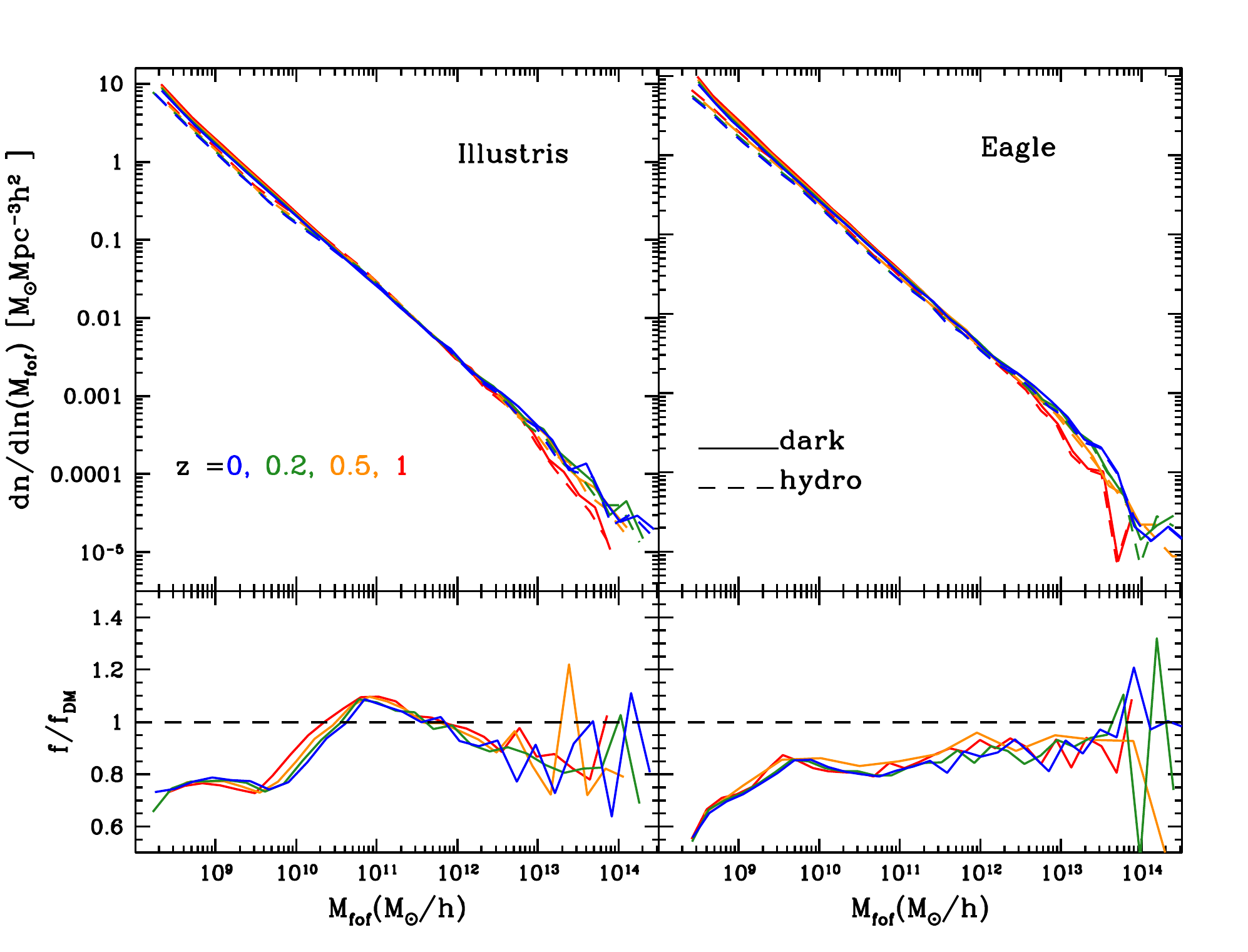}
\includegraphics[width=0.7\hsize]{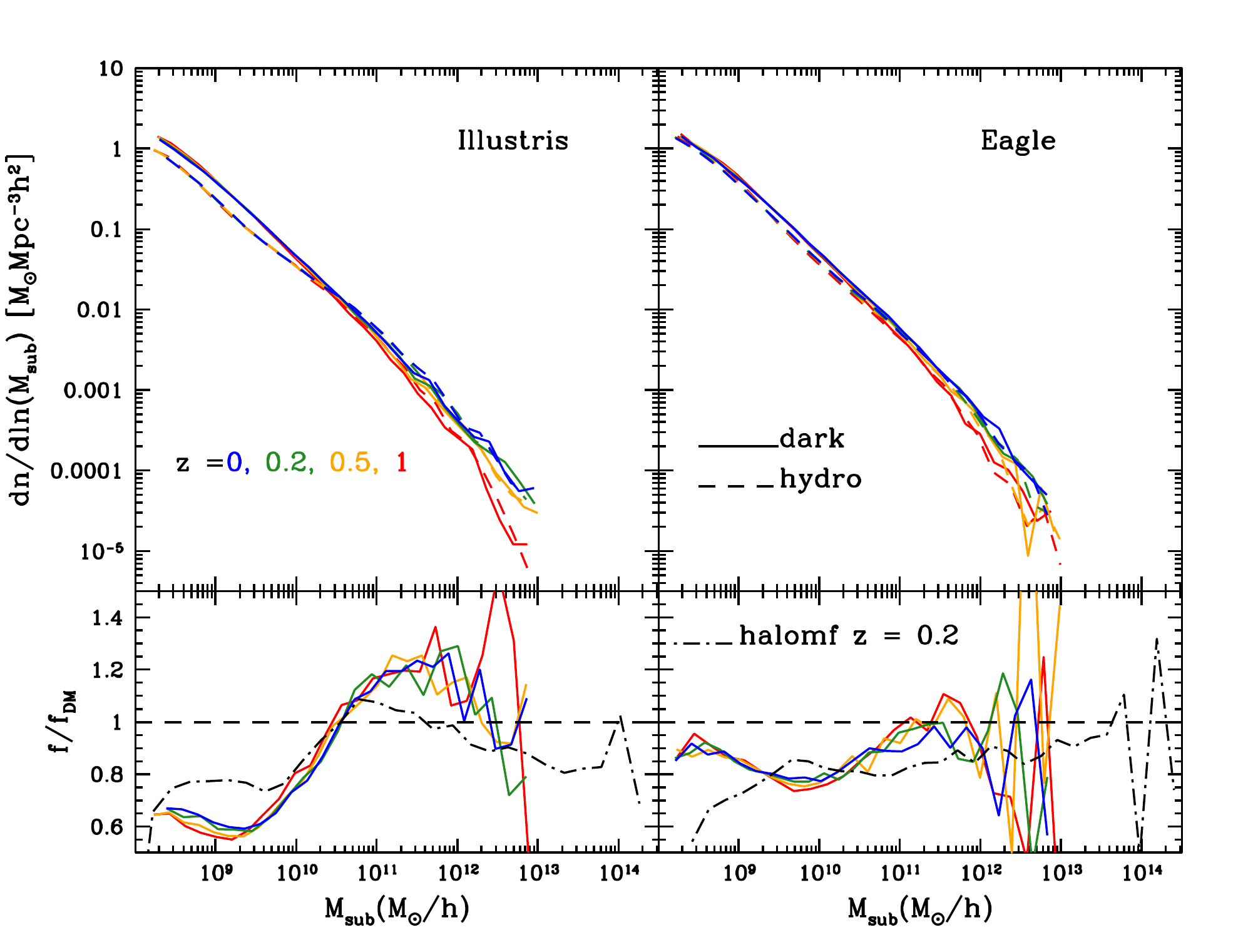}
\caption{Halo and (global) subhalo mass  function in the Illustris and
  EAGLE simulations - top and  bottom panels respectively; the results
  from the dark-matter-only and  the full hydrodynamical runs are show
  respectively by  solid and dashed  lines, different  colors stand
  for different  redshifts (z=0,0.2,0.5,1).  The lower  panel show the
  fraction of haloes or subhaloes found in the hydro run, with respect
  to the  dark-matter-only one.  The ratio of the  halo mass functions
  reflect    those    already    found    in    \citet{vogel14}    and
  \citet{schaller15}.\label{massfunc}}
\end{figure*}

First of  all, we want  to investigate how the  differences in
  the subhalo  population can  be related  to those  in the  halo mass
  function: a  different number of  haloes available for  accretion at
  redshift $z_{j}$ would  induce a different subhalo  mass function at
  redshift  $z_{k}<z_{j}$.  Figure  \ref{massfunc}  shows  the  halo
(top  panels)  and   subhalo  (bottom  panels)  mass
function for both  simulations, at four redshifts between 0  and 1. In
the upper panel  we plot the halo mass function,  using all the haloes
in the FOF  catalogues and, thus, in this case  we choose $M_{fof}$ as
halo mass;  results from the dark-matter-only  and full hydrodynamical
run are  represented, respectively, with  solid and dashed  lines.  In
the small  lower panels, we  highlight the difference between  the two
cases: by looking at the  fractional difference of halo counts between
the hydro and the dark-matter-only run, we notice  some similarities
and some differences between  the two simulations.  We observe
  the predicted reduction of the number of objects from the DMO to the
  hydrodynamical  runs, but  while in  the EAGLE  case, this  lack of
structures is maintained at all  masses (even the AGN feedback
  cannot expel enough baryons to reduce the halo mass at the high-mass
  end, so that the ratio tends to 1, as in \citealt{schaller15}), in the
Illustris the  reduction is not  constant and there  are more
intermediate  mass haloes  in the  hydro run  than in  the DMO
  one.  This  behaviour can  be explained as  the combined  effect of
stellar  and  AGN  feedback:  they  both  are  less  efficient  around
$10^{11}M_{\odot}h^{-1}$,  leaving   these  masses  unaltered   and  so
increasing their  abundance due to  the contribution of  higher masses
that are shifted to these lower values. This is consistent with what
  is shown in \citet{vogel14}; indeed \citet{cui12,cui14} suggest
  that neglecting  the AGN feedback  can lead  to an increase  of the
halo mass function at the massive end.

The second  panel of  Figure \ref{massfunc}  shows the  global subhalo
mass function, in the same units  of the halo mass function. The colour
scheme is  the same; the  dot-dashed black  curve in the  lower panels
shows  the fractional change  of  the  halo mass  function  at  $z=0.2$ from  the
previous panel,  for a more  straightforward comparison.  We  notice a
clear relation  between the two  mass functions: a considerable  part of
the  difference   in  the  subhalo   counts  between  the   hydro  and
dark-matter-only run,  can be attributed to  the underlying difference
in the halo  mass function.  Having less small structures  that can be
accreted by larger haloes at any  redshift leads to a different number
of subhaloes - also enhanced by the fact that not all the small haloes
at $z_{i}<  z$ will be  accreted.  Nevertheless, the residuals  do not
correspond exactly,  and the additional differences  can be attributed
to the action  of baryonic physics, as  cooling, adiabatic contraction
and tidal forces inside the halo.

\subsection{Mass bins}

We    select     the    dark    matter    haloes     with    a    mass
$M_{200} = [10^{12.5},  10^{14}]M_{\odot}h^{-1}$, consistent with the
mass range of  massive elliptical galaxies with which  we will compare
further on  in this  work.  In this  mass range,  the baryonic
  effect on the halo mass function is similar for the two simulations,
  leading to a $\simeq 10\%$ reduction  in the number of haloes.  The
minimum  subhalo   mass  that  we   show  in  our   plots  is
$M\simeq 2\times  10^{8} M_{\odot} h^{-1}$,  which corresponds
to $n_{part}\simeq 20-30$ depending on  the run.  Previous works showed
that the subhalo identification may not be reliable with less than 100
particles    \citep{onions12},     which    would     correspond    to
$M>10^{9}  M_{\odot}  h^{-1}$.   Here,  we keep  the  small  subhaloes
in the  plots, but we remind  that the results for  this mass
bin must  be interpreted with  caution. We use  only subhaloes
  with more than 100 particles to fit the mass function.

\begin{figure}
\includegraphics[width=1.03\hsize]{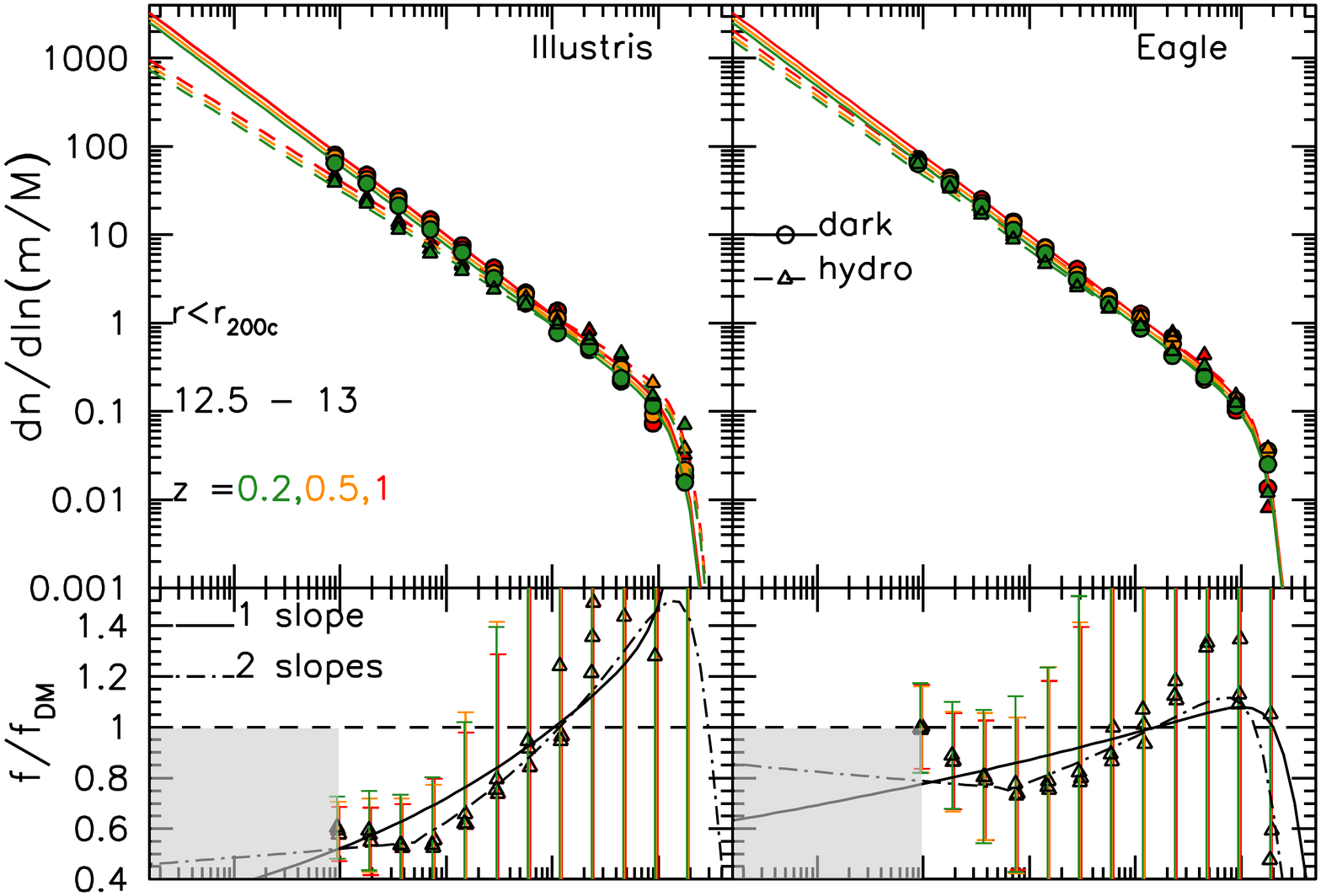}
\includegraphics[width=1.03\hsize]{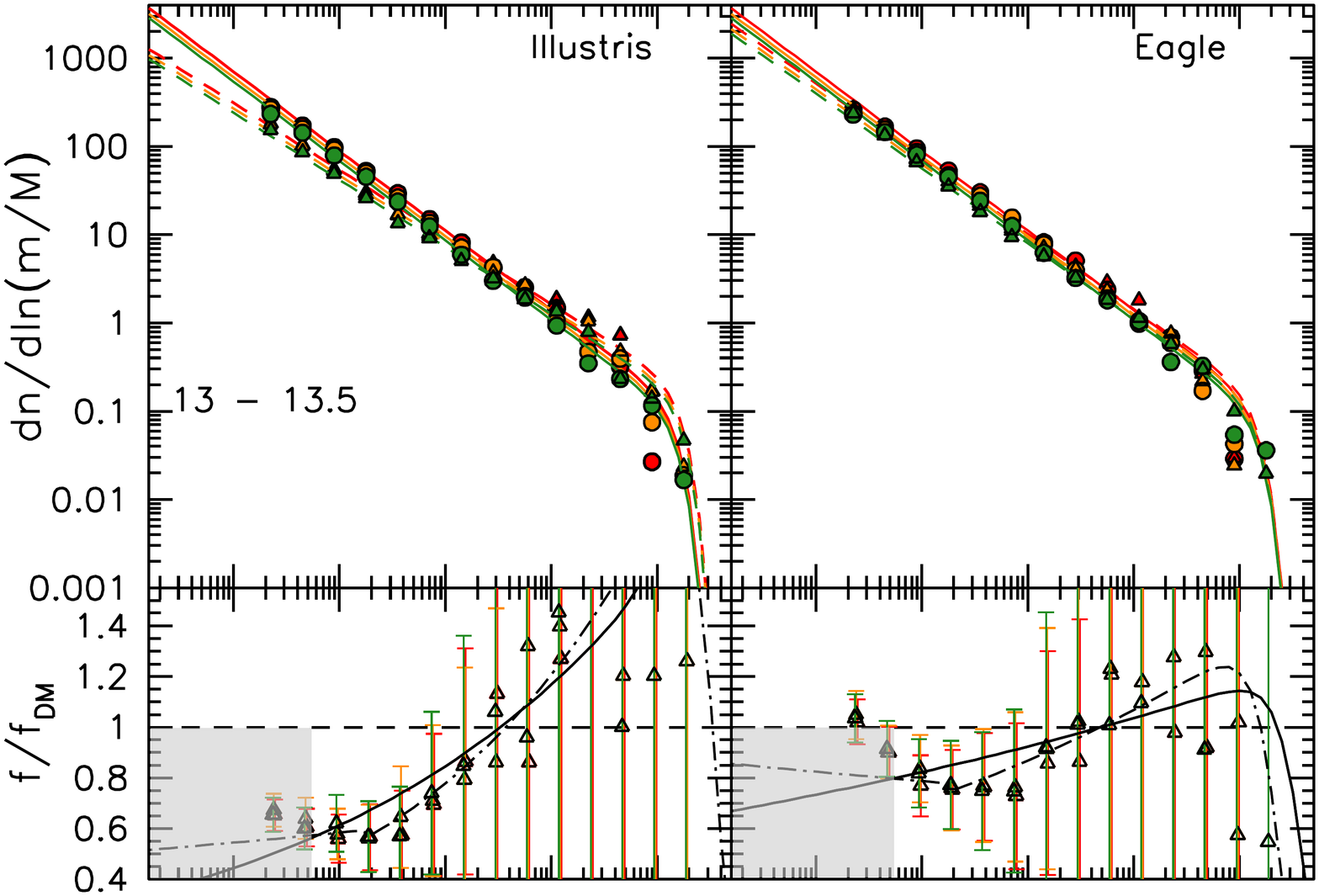}
\includegraphics[width=1.03\hsize]{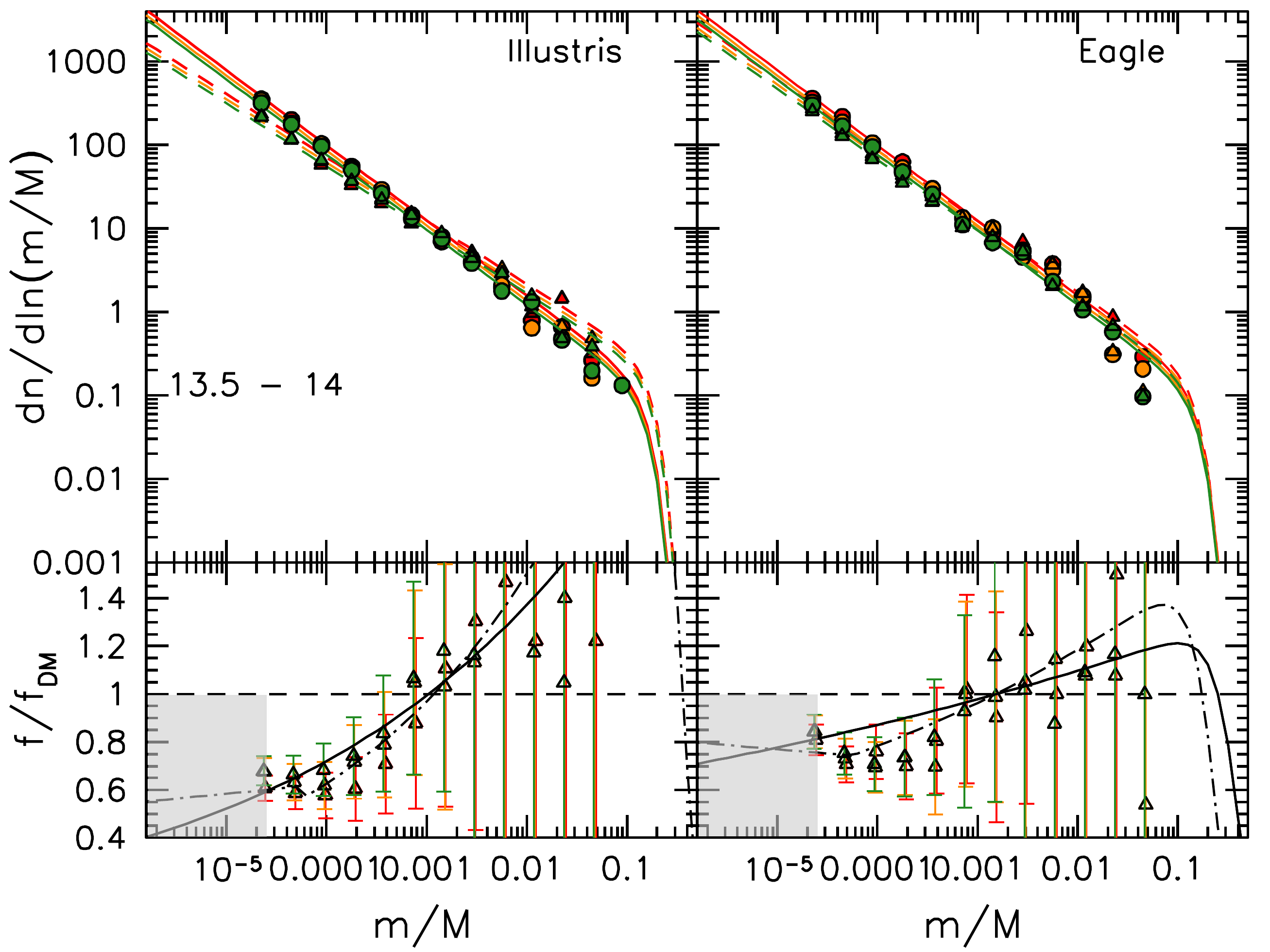}
\caption{Subhalo  mass function  for three  bins in  halo mass  in our
  sample ($m/M=m(subhalo)/M_{200}$).  The colour scheme is the same as
  in Figure \ref{massfunc}.  \emph{Main panels}: circles and triangles
  show our measurements for the  dark and hydro simulations, fitted by
  the curves  coming from Equation  \ref{fit} (again solid  and dashed
  lines).  \emph{Inset panels}: fraction  of haloes or subhaloes found
  in the hydro  run, with respect to  the dark matter only  one - with
  Poissonian  errors.   While  the  points  stand  for  the  fractions
  calculated  from the  data, the  solid  black line  shows the  ratio
  between the standard fitting functions: as  they scale only with redshift for
  a given  mass bin,  this is  the same for  all the  three considered
  redshifts.  The dot-dashed line shows the same, but in this
    case the hydro mass function is modelled separately for low and
    high mass subhaloes and thus has two different slopes (see Table \ref{tab_sim1}).
The  gray region correspond to subhaloes  with less than
  100 particles. \label{massfunc2}}
\end{figure}

Figure \ref{massfunc2} shows  the subhalo mass function  in three bins
of halo mass, for the two simulations and both the dark-mater-only and
hydro   run.   We   fit   our  measurements   with   the  model   from
\citet{giocoli10a}:
\begin{equation}\label{fit}
\frac{1}{M}\frac{dn(z)}{d\ln m}=(1+z)^{1/2}A_{M}m^{\alpha}\exp\left[-\beta\left(\frac{m}{M}\right)^{3}\right],
\end{equation}
where the parameters $(\alpha,\beta,A_{M})$ are respectively the slope
(assumed   negative),  the   exponential   cut-off  and   the
normalization of the  curve. We fit the  subhalo mass function
  corresponding to different  halo masses and redshifts  (as in Figure
  \ref{massfunc2}) and  then take  the average value  of the  best fit
  slope. Consistently  with previous works, for  the dark-matter only
case we recover the standard  value of $\alpha=-0.9$; since the values
in \citet{giocoli10a} are  for virial haloes, the  other parameters of
our best-fit  differ.  For  the runs with  baryons, we  find a
  best fit  slope which  is less  steep than the  DMO case  (see Table
  \ref{tab_sim1})  .  Note  that  the fitting  functions are  shifted
vertically by the redshift dependence,  and thus the residuals between
them shift  horizontally in  the residual  panels, fitting  the points
well  within the  errorbars.  As  can be  understood from  the
  lower  panels of  Figures  \ref{massfunc}  and \ref{massfunc2},  the
  presence of baryons modifies the shape of the subhalo mass function.
  In order to provide a simple description and be able to compare with
  observational results \citep{vegetti14}, we fit Equation  \ref{fit} both to the DMO and
  the hydro  runs. However it  should be noted  that for the  latter a
  single power-law model may not be as accurate for the entire subhalo
  mass range as for the DMO case. For this reason, we also calculate
  the best-fit slope separately for two subhalo mass intervals and the
best fit values are listed in the third and fourth columns of Table
\ref{tab_sim1}. We see that the mass function is steeper for the
low-mass subhaloes (when $log(M)<9.5$, which corresponds to the
plateau in the residual panels of Figure \ref{massfunc2}) and has a
slope similar to the DMO case, consistently with what found in
\citet{sawala17}.  The dot-dashed line in the lower panels of Figure
\ref{massfunc2} shows the ratio between the DMO subhalo mass function
and the hydro one which combines the two slopes.

\begin{table} \centering
\begin{tabular}{|c|c|c|c|}   \hline 
\multicolumn{2}{c}{Subhalo mass function slope}  \\ \hline  
sim & $\alpha$ & $\alpha(log(M)<9.5)$ & $\alpha(log(M)>9.5)$\\  
\hline 
DMO & -0.90 $\pm$ 0.03 & - & -\\
EAGLE & -0.85 $\pm$ 0.04  & -0.91 $\pm$ 0.03 &  -0.82 $\pm$ 0.06\\
Illustris & -0.76 $\pm$ 0.02  & -0.87 $\pm$ 0.03 &  -0.72 $\pm$ 0.09 \\
 \hline
\end{tabular}
\caption{Column 2: best fit slope of the subhalo mass function from
  Equation (\ref{fit}). Column 3-4: best fit slopes obtained
    modelling the subhalo mass function separately at the low and high
  mass end. \label{tab_sim1}}
\end{table}

Different colors  stand for different  redshifts and the  solid black
line shows the  ratio between the fitting functions.   To test whether
the residuals  depend on  the halo  mass definition,  we re-calculated
them also for the whole FOF halo.  They behave consistently with those
within  $r_{200}$, showing  that the  lack of  small subhaloes  in the
simulations  does not  depend only  on the  distance from  the centre,
supporting the point that the variation  in the halo mass function may
dominate over a faster disruption rate inside the halo.

Looking at  the errorbars we  notice that the difference  in fractions
between the  two simulations is significant  only at the low  mass end
where the number of objects is high enough and therefore the errorbars
are smaller.   We will explore  the low mass  end of the  subhalo mass
function with higher  resolution simulations in a  follow-up paper: in
order  to distinguish  between the  effect of  baryons and  those from
different models of  dark matter (such as warm dark  matter) we need a
good statistic at $10^{6}-10^{7}M_{\odot}$.

\section{Subhalo counts} \label{sec_count} In this section we present
and model the distribution of subhaloes as a function of radius, both
in terms of their three dimensional number density and in
projection. As we adopted $M_{200}$ as halo mass, we use $r_{200}$
accordingly.

\begin{figure}
\includegraphics[width=\hsize]{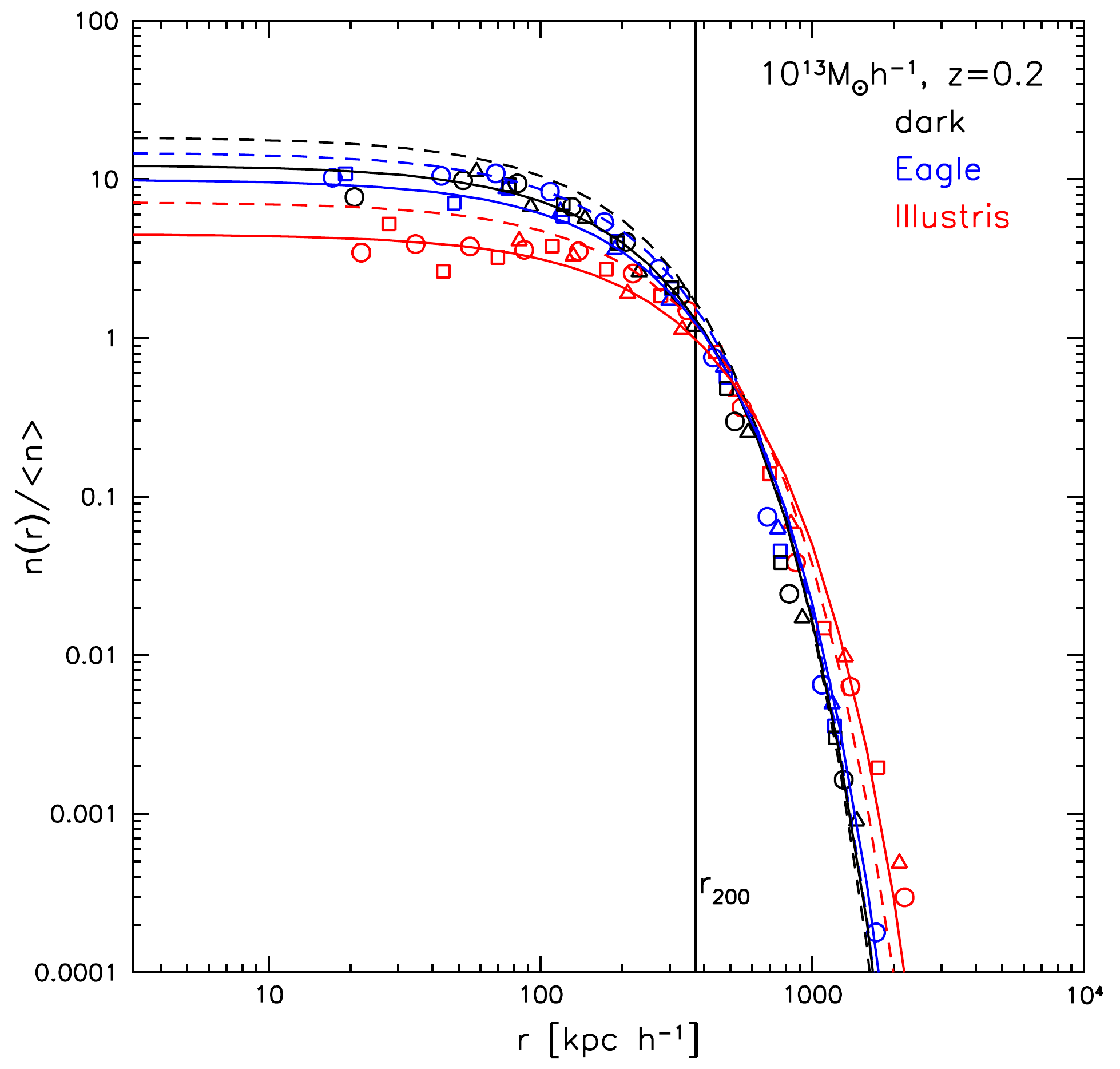}
\caption{Number  of  subhaloes  as  a   function  of  radius.   As  in
  \citet{springel08b},  the radial  distribution of  subhaloes can  be
  well fitted  by an Einasto  profile. Different point  types indicate
  three  bins  of  subhalo   masses  -  $10^{8}$  (circles),  $10^{9}$
  (squares)  and   $10^{10}M_{\odot}h^{-1}$  (triangles):   all  these
  subhaloes lie  on the same curve,  when the number is  normalized by
  the  mean number  density of  subhaloes  of that  specific mass  bin
  within $r_{200}$. Solid  lines show the best fit  Einasto profile to
  the $z=0.2$ points, while the same  fit for $z=0.5$ (for which we do
  not show the points) are given  by the dashed lines. Other mass bins
  give similar results,  and we provide the parameters  of the Einasto
  fits in Table \ref{tab_sim2}.\label{nsub}}
\end{figure}

\begin{table} \centering
\begin{tabular}{|c|c|c|c|}   \hline 
\multicolumn{4}{c}{number density of subhaloes}\\
\hline
sim & $10^{13}$ &  $10^{13.5}$ &  $10^{14}$\\  
\hline 
\multicolumn{4}{c}{z = 0.2}  \\ \hline  

DMO & (1.79 , 309)& (2.12 , 436) & (1.12 , 881)\\
EAGLE & (1.62 , 330)&(1.92 , 447)&(1.04 , 895)\\
Illustris&(0.732 , 438)&(1.04 , 488)&(1.11 , 760)\\
 \hline
\multicolumn{4}{c}{z = 0.5}  \\ \hline  
DMO & (2.74 , 291)& (3.23 , 388) & (1.17 , 381)\\
EAGLE & (2.4 , 306)&(4.38 , 360)&(1.61 , 514)\\
Illustris&(0.732 , 438)&(2.02 , 733)&(2.02 , 677)\\
 \hline
\end{tabular}
\caption{Best fit parameters of the Einasto fit to the number density of
  subhaloes as a function of radius. For each
  redshift-mass-simulation combination, we give the best fitting values of
  $(\rho,r_{s})$ of the fit; we find that in all cases $\alpha=1.1$
  works well. Some of the fitting functions are shown in Figure
  \ref{nsub}, together with the points.\label{tab_sim2}}
\end{table}

It has been shown in previous works (e.g. \citealt{springel08a}) that the number
of  subhaloes as  a function  of  radius can  be well fit  by an  Einasto
profile:
\begin{equation}
\rho(r) = \rho_{s}\exp\left\{-\frac{2}{\alpha}\left[\left(\frac{r}{r_{s}}\right)^{\alpha}-1\right]\right\}.
\end{equation} 

In Figure  \ref{nsub} we  show the  radial distribution  of subhaloes:
for  each simulation,  the  distribution of  subhaloes can  be
  modelled by  the same curve, when  the number is normalized  by the
mean  number density  of subhaloes  in that  specific mass  bin within
$r_{200}$.   Solid lines  show the  best  fit Einasto  profile to  the
$z=0.2$ points,  while the same fit  for $z=0.5$ (for which  we do not
show the points) are given by the  dashed lines. In all cases, we find
that  our  points  are  well   fitted  by  an  Einasto  profiles  with
$\alpha=1.1$, while  the values of normalization  $\rho_{s}$ and scale
radius  $r_{s}$   vary  with  halo   mass  and  redshift.    In  Table
\ref{tab_sim2} we provide the best-fit  values for each combination of
redshift, mass bin and simulation.   At fixed redshift, increasing the
halo mass corresponds to an horizontal shift in the profiles (i.e.  an
increased number density of subhaloes at large distances); at
fixed mass, increasing  $z$ leads to an increase  in the normalization
(i.e.  an overall increased number density of subhaloes).  We
also note  that for  high halo  masses, the  three profiles  have more
similar  shapes,  indicating  that  the  effect  of  baryons  is  less
important in this mass range.

\begin{figure*}
\includegraphics[width=0.4\hsize]{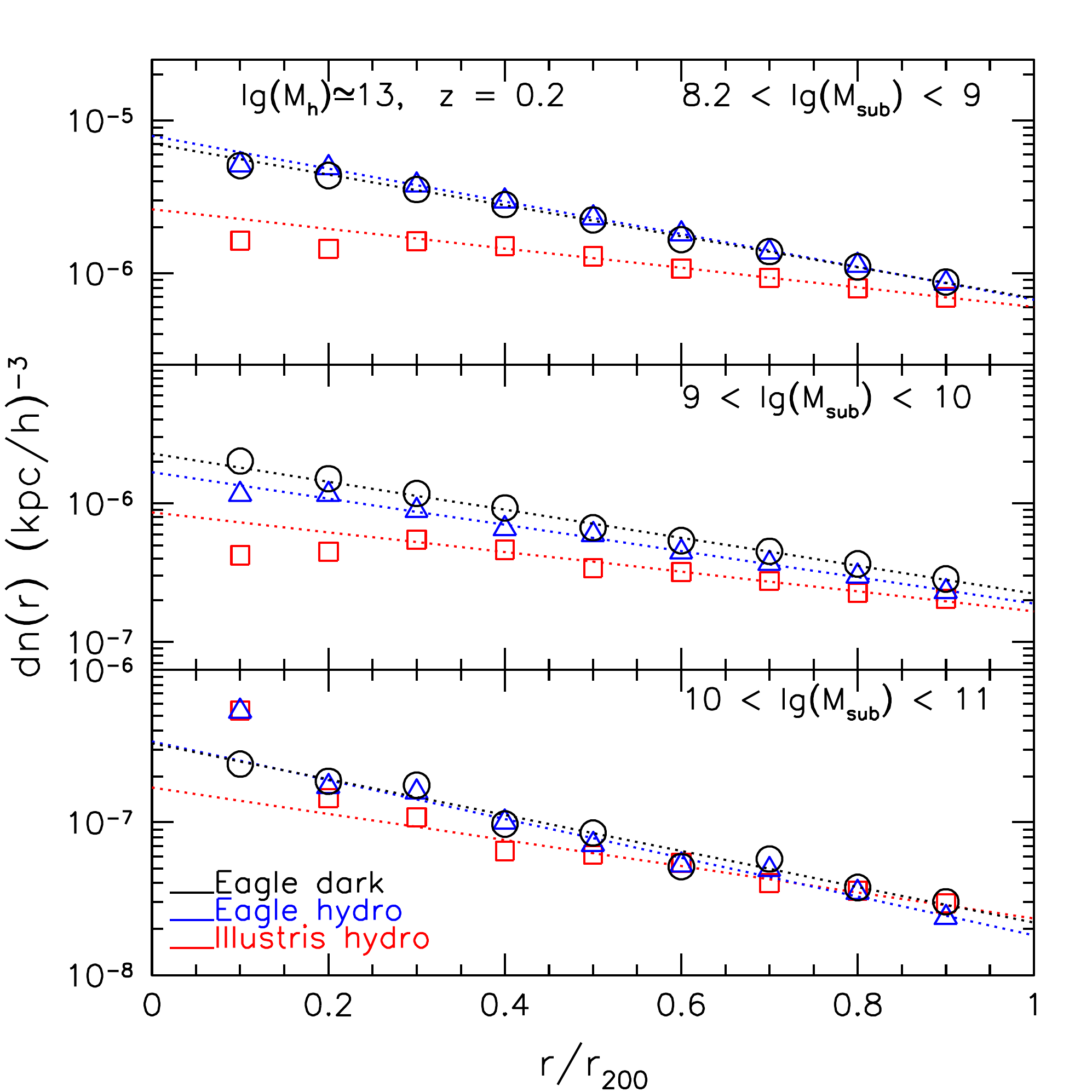}
\includegraphics[width=0.4\hsize]{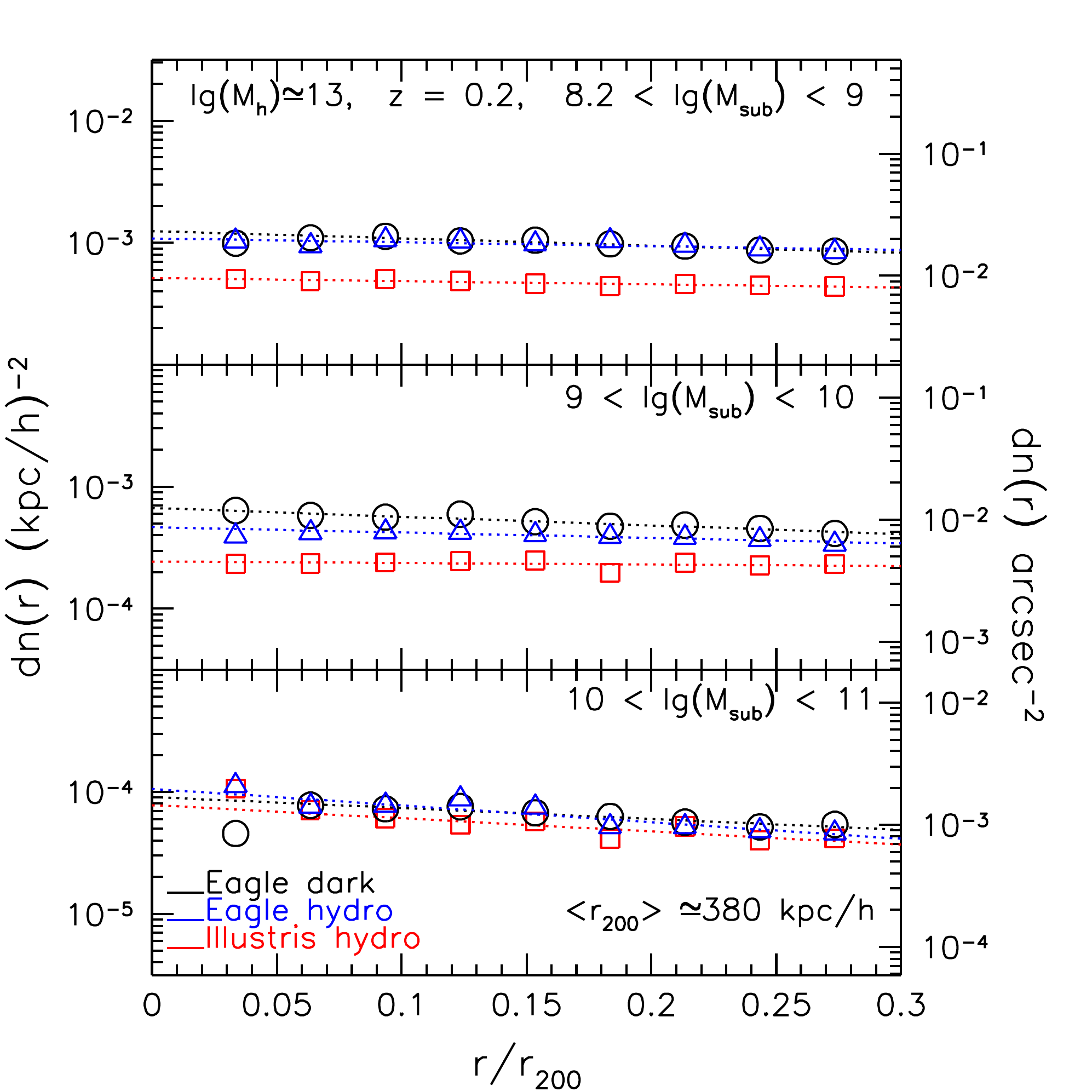}
\includegraphics[width=0.4\hsize]{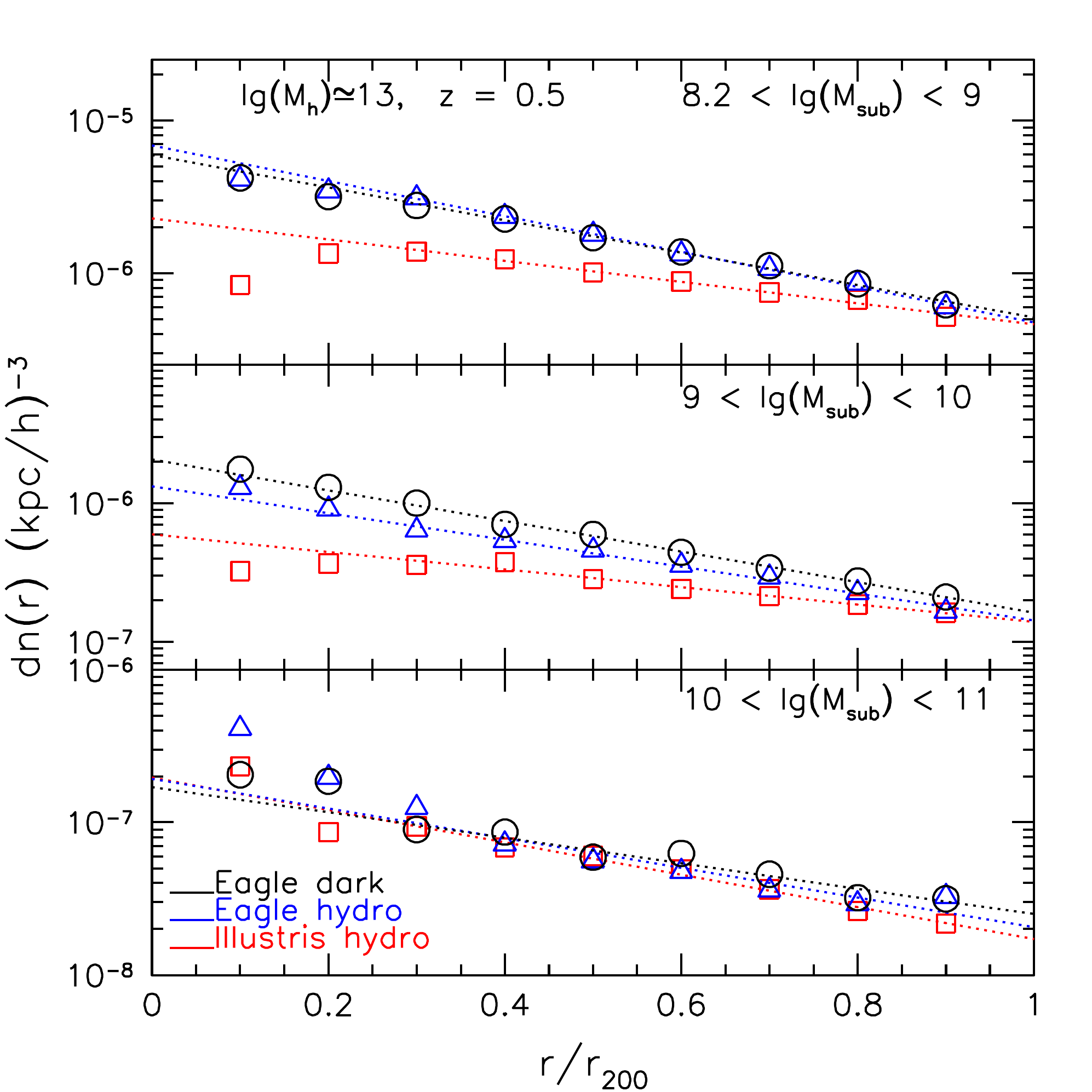}
\includegraphics[width=0.4\hsize]{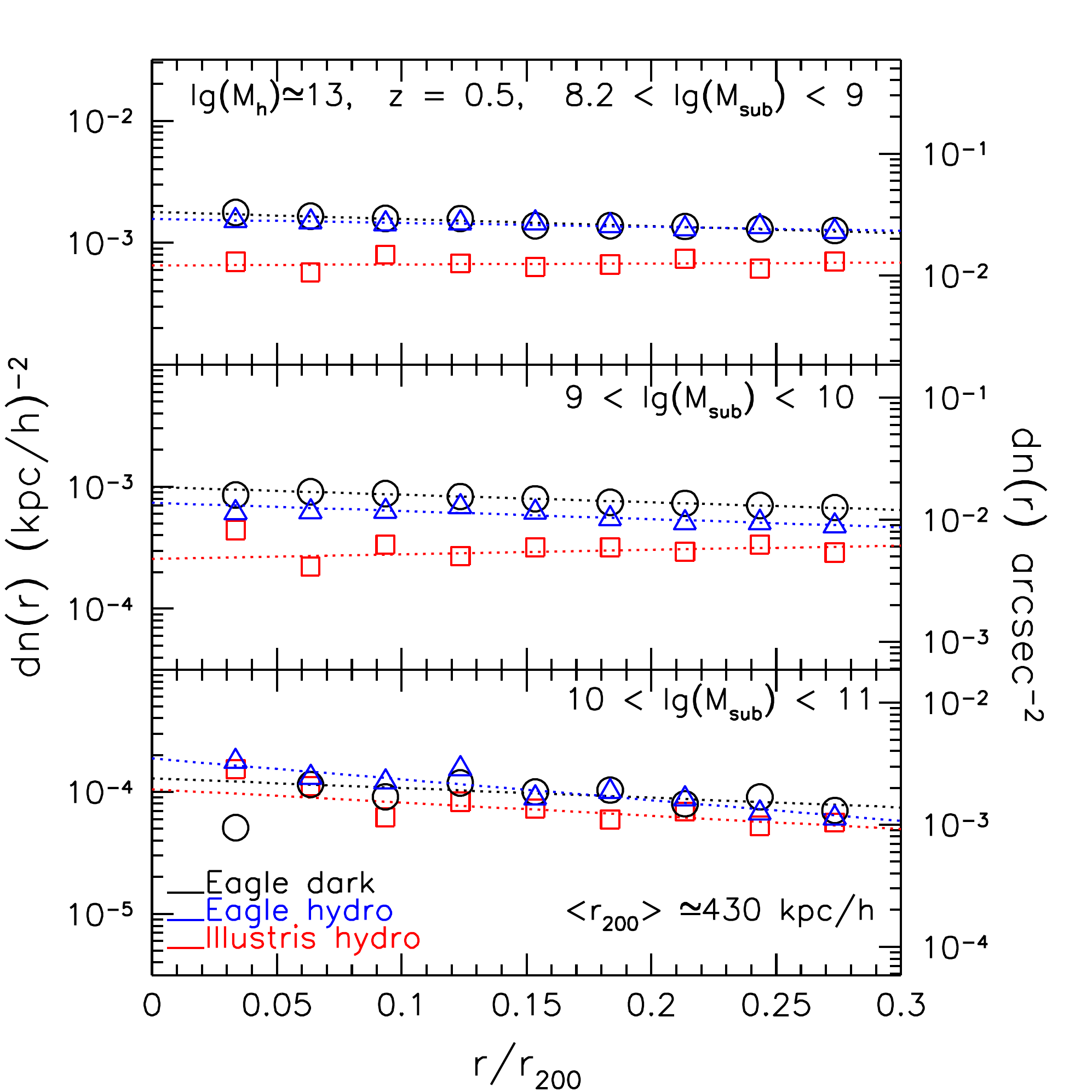}
\caption{Subhalo counts at redshift $z=0.2$ and $z=0.5$.  Each
    horizontal  panel  corresponds   to  a  subhalo  mass   bin  (in
    $M_{\odot} h^{-1}$). $(Left)$ Average  3D number of subhaloes, in
  units  of physical  $(kpc/h)^{3}$.   $(Right)$  projected number  of
  subhaloes  (averaged  over  three  projections per  halo),  in  units  of
  physical $(kpc/h)^{2}$ or $arcsec^{2}$. In all the panels, we divide
  the subhaloes in  three mass bins, to show how  the abundance scales
  approximately  one order of  magnitudes for each decade  in mass.
  Results for the DMO, EAGLE  and Illustris hydro runs are represented
  respectively by black  circles, blue triangles and  red squared; the
  dotted   lines    are   linear    best   fit   relations    to   the
  points. \label{ndensity}}
\end{figure*}

Figure \ref{ndensity}  shows the  average subhalo  counts in  units of
$(kpc/h)^{-3}$,  $(kpc/h)^{-2}$ and  $arcsec^{-2}$. as  a function  of
distance from  the centre in three  and two dimensions, for  haloes of
mass $M_{200}\simeq  10^{13}M_{\odot}h^{-1}$ at redshift 0.2  and 0.5;
the subhalo  counts are  averaged over all  the haloes  in the
  selected mass bin and over three  projections per system for the two
  dimensional measurements. We  compare the dark-matter-only results,
the EAGLE (blue  open triangles) and the Illustris  (red open squares)
hydrodynamical  simulations.  We consider  three  logarithmic
  bins  in subhalo  mass  (in $M_{\odot}h^{-1}$:  (8.2-9), (9-10)  and
  (10-11).   The  left  panels  show the  average  three  dimensional
subhalo counts within $r_{200}$, expressed  as number of subhaloes per
$(kpc/h)^{-3}$.   The  general  trend  is very  similar  for  all  the
simulations, but we note again a different effect of the two models of
baryonic physics: in  the Illustris simulation, the  lack of subhaloes
is again more evident than in EAGLE, reflecting what could be inferred
from the  subhalo mass  functions.  In  projection (right  panel), the
distributions flatten and the relative behaviour of the three cases is
the same, similarly to what found in \citet{xuD15} - this time we plot
only the subhalo  counts within $0.3\times r_{200}$,  because only the
central parts  of the halo  are relevant for lensing.   Projecting all
the  substructures inside  the FOF  group gives  a slightly  different
results in the higher subhalo mass  bin; this is explained by the fact
that bigger subhaloes are found in  the outer region of haloes and the
FOF group is more extended than $r_{200}$.

\section{A comparison with observations} \label{sec_obs}

\subsection{The SLACS survey}
The  Sloan  Lens ACS  (SLACS)  Survey  is  an efficient  Hubble  Space
Telescope (HST)  Snapshot imaging survey for  galaxy-scale strong
gravitational lenses  \citep{bolton06}. The SLACS survey  was optimized
to detect bright early-type lens galaxies with faint lensed sources in
order to  increase the sample  of known gravitational  lenses suitable
for detailed  lensing, photometric, and dynamical  modeling. SLACS has
identified nearly 100 lenses and lens candidates, which have a stellar
mass ranging from  $10^{10.5}$ to $10^{11.8}M_{\odot}$ \citep{auger09}
and  estimated  total  masses   of  the  order  of  $10^{13}M_{\odot}$
\citep{auger10a}, at  redshift $z\simeq 0.2$ -  spanning approximately
$0.06\leq     z_{lens}\leq     0.5$     for     the     lenses     and
$0.2\leq z_{lens}\leq  1.3$ for  the source.  They can  be considered
representative of  the population of massive  early-type galaxies with
stellar masses $M_{*}\geq 10^{11}M_{\odot}$ \citep{auger09}.

In    this   paper,    we    make   use    of    the   results    from
\citet{auger09,auger10a,auger10b},  in  order   to  select  SLACS-like
candidates  in   the  EAGLE   and  Illustris  hydro   simulations.

\subsection{Considerations on the IMF}

Even  though  it  could   seem  more  straightforward,  selecting  the
simulated galaxies in terms of the total stellar mass may present some
problems concerning  the chosen  IMF: the  SLACS total  stellar masses
have been calculated  by \citet{auger09}, using both  the Chabrier and
the Salpeter  IMF and the  latter has been  shown to be  the preferred
model.   \citet{auger10a}  ruled out  a  Chabrier  mass function  when
calculating the total halo mass (this is also supported by the
  findings of  \citet{grillo09}.   On the other
hand, both  the EAGLE and Illustris  simulations have been run  with a
Chabrier  IMF  and   rescaling  the  mass  with   the  usual  relation
$M_{Sa} =  M_{Ch}/0.55$ would  not yield  to a  meaningful comparison:
changing the IMF in the  simulation code would require modification in
the baryon and subgrid models, leading to more complicated differences
than a simple rescaling.  For this reason, following \citet{xuD16}, we
avoid a  selection by total stellar  mass and prefer to  use dynamical
measurements and  velocity dispersion instead.  This  presents another
advantage: as can  be seen in Figure \ref{mstar}, the  stellar mass of
the central galaxy is on average  different in the two simulations for
a halo  of a  given total  mass and so  using it  as a  main selection
criterion could enhance  the difference in total mass  between the two
samples.

\subsection{Selection of SLACS analogues}

\begin{figure}
\includegraphics[width=\hsize]{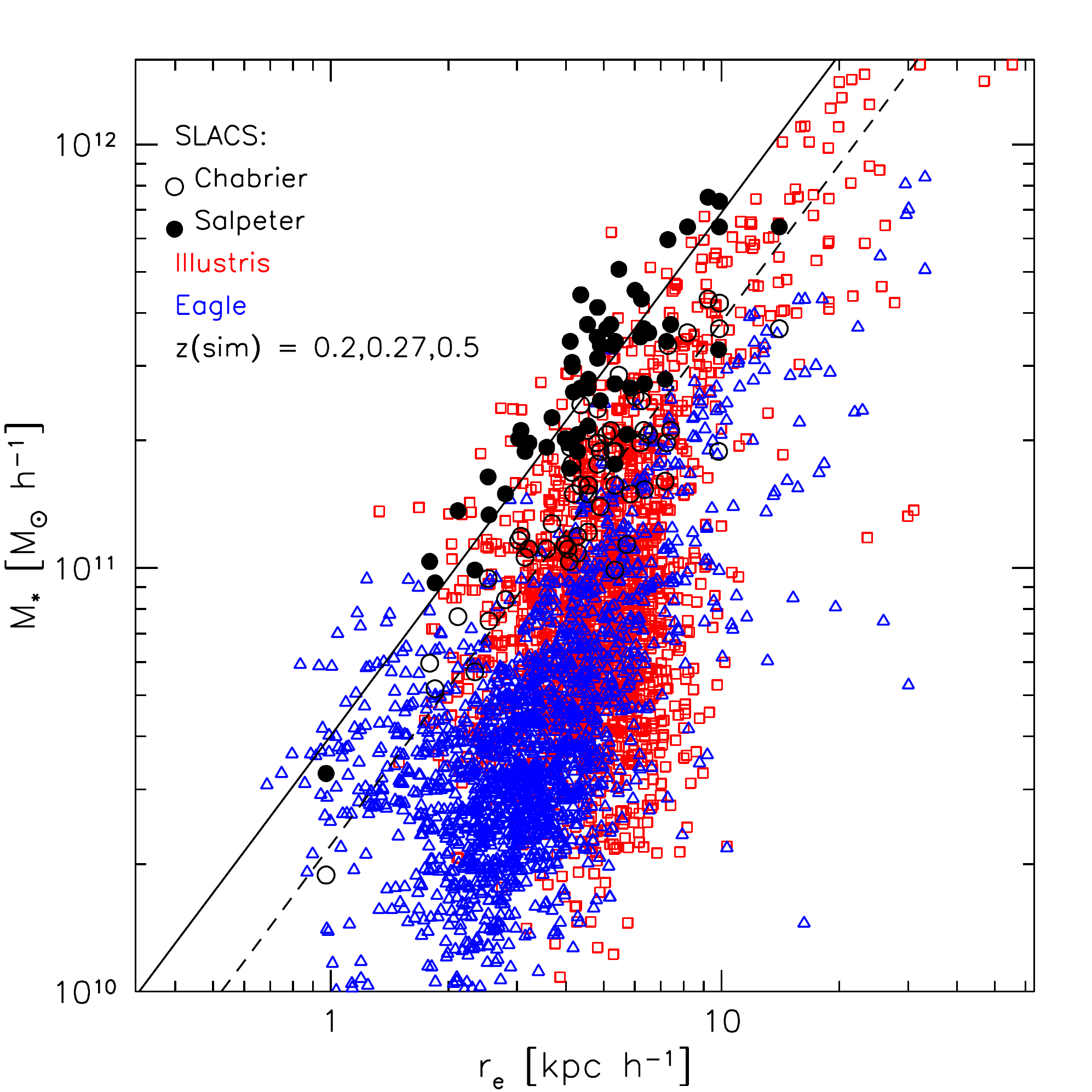}
\includegraphics[width=\hsize]{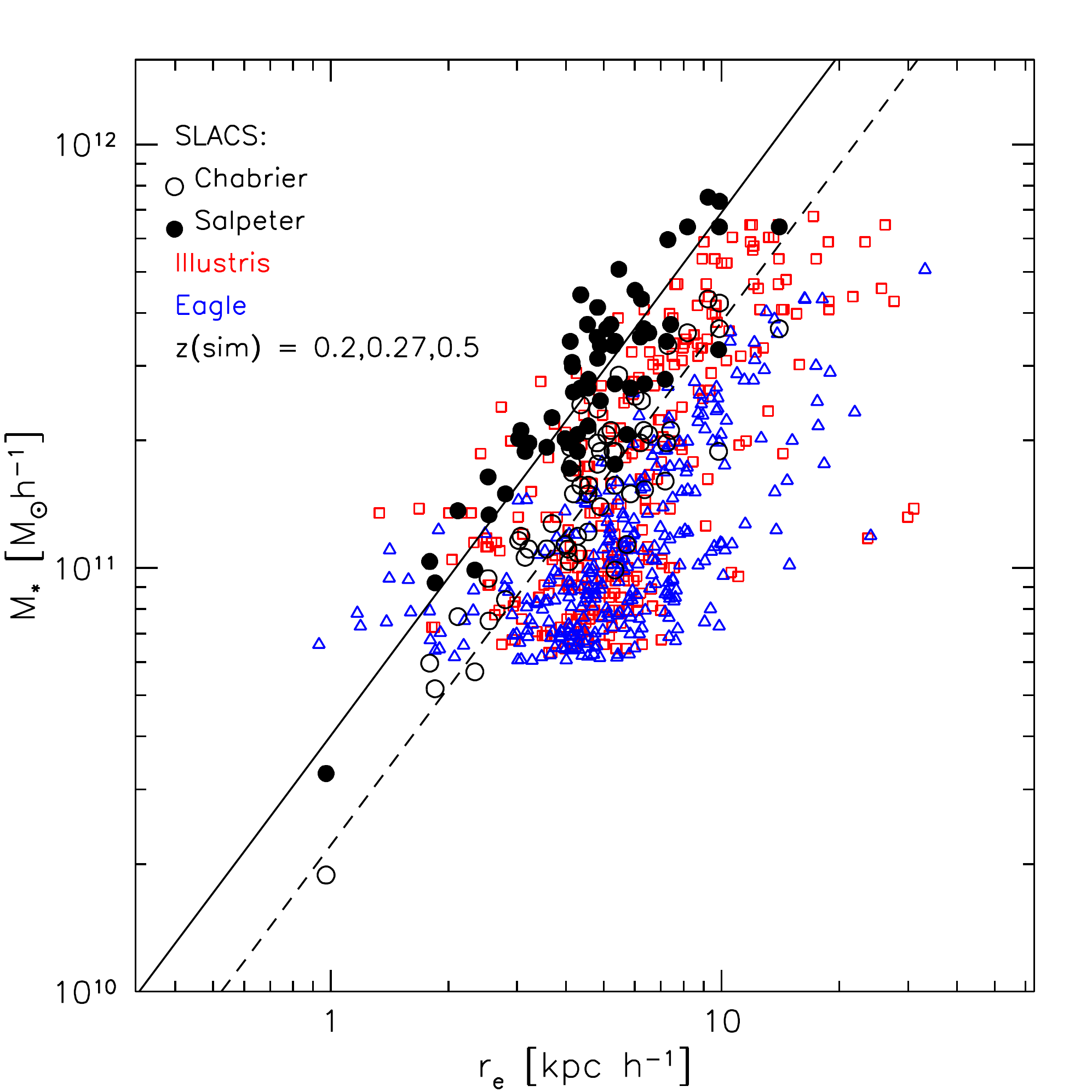}
\caption{$Top$: Total stellar mass vs. effective radius, for the SLACS
  lenses (black filled circles for a Salpeter IMF and open circles for
  a Chabrier IMF) and the galaxies  in the EAGLE (blue open triangles)
  and  Illustris  (red open  squares)  simulations.   The black  solid
  (dashed) line  shows the best fit  to the SLACS data  for a Salpeter
  (Chabrier) IMF by \citet{auger10b}.   Here we show all the galaxies
  that lie in the  $M_{200}=10^{12}-10^{14} M_{\odot}h^{-1}$ mass bin.
  $Bottom$:  same   for  the   galaxies  selected   through  dynamical
  properties and velocity dispersion ($\sigma_{v}=160-400 km/s$).
  \label{slacs}}
\end{figure}

In order to  select analogues of the SLACS lenses  from the considered
simulations,  we  start by  excluding  all  haloes which  are  clearly
unrelaxed: we  exclude those  for which the  distance between  the halo
centre of  mass and the  position of the  minimum of the  potential is
more than  5$\%$ of the  virial radius,  as they may  include multiple
components    and    are    not    suitable    for    our    selection
\citep{neto07,ludlow12}.The  top panel of  Figure \ref{slacs}
  shows    all     galaxies    selected    only    by     halo    mass
  $M_{200}=10^{12}-10^{14} M_{\odot}h^{-1}$ and  relaxation; unlike in
  previous sections,  here we  also include  haloes with  masses lower
  than $10^{12.5}  M_{\odot}h^{-1}$, in order to  reproduce the virial
  mass range of  the SLACS lenses The SLACS lenses  are represented by
  the black circles  (open for Chabrier and filled  for Salpeter IMF),
  while the simulated  galaxies are represented in  red (open squares)
  and blue (open  triangles), for the Illustris  and EAGLE simulations
  respectively.  These come from three  snapshots of the simulations -
  the closest to  the $z$ range of  the SLACS and the  closest to each
  other.  As has been found  in \citet{auger10b}, the SLACS lenses are
  well  fit  by   a  linear  relation  in   the  $M_{*}-r_{e}$  space,
  represented by the dashed and solid black lines for the two IMFs. 

We then apply a dynamical selection, identifying the galaxies that are
most probably ellipticals, or at  least bulge dominated.  This kind of
information comes from dynamical measures already present in the EAGLE
and  Illustris catalogues,  which allow  to distinguish  the disk  and
bulge components and give very  similar informations and results.  

For the EAGLE run, we  define galaxies with a counter-rotating
  stellar fraction  inside 20 kpc of  at least 25\% to  be elliptical.
For Illustris, the selection is based on the specific angular momentum
\citep{teklu15,snyder15} through  the parameter $\epsilon=J_{z}/J(E)$,
where $J_{z}$ is  the specific angular momentum  and $J(E)$ is
  the maximum local angular momentum  of the stellar particles.  This
quantity has been calculated for each stellar particle and we know:
\begin{itemize}
\item the fraction of stars with $\epsilon>0.7$, which is a common
  definition of the disk stars - those with significant (positive)
  rotational support
\item the fraction of stars with $\epsilon<0$, multiplied by two, which
  in turn commonly defines the bulge.
\end{itemize}
Following \citet{teklu15}, the  galaxy is disk dominated  if the first
quantity is greater than 0.4 and  bulge dominated if the second one is
larger than  0.6. Our selection  combine these two criteria,  with the
second one  being much more  restrictive than the first.   Despite the
fact  that  the quantities  calculated  for  the two  simulations  are
different, the resulting selection is  similar: both criteria allow us
to estimate the mass  of the disk/bulge of the galaxy  and to rule out
systems which clearly have a disk works well.

Among  the galaxies  identified by  the dynamical  selection, we  then
choose galaxies that have a velocity dispersion similar to that of the
SLACS lenses  (160-400 km/s) . This  is calculated within half  of the
effective radius  $r_{e}$, which  traces half of  the light.   For the
simulated galaxies, we calculate a  projected half mass radius, using
the central  galaxies corresponding to our  selected haloes, averaging
over the three projections along the  axes.  This quantity should be a
good  counterpart of  the  effective  radius and  we  use  it also  to
recalculate the velocity dispersion.

The bottom panel of Figure \ref{slacs} illustrates the result
of our  selection in the $r_{e}$  - $M_{*}$ space, showing  the sample
selected using  dynamical properties and velocity  dispersion, proving
that we are able to choose  in simulations objects very similar to the
SLACS  lenses.   The main  restrictions  are  given by  the  dynamical
criteria:  this reinforces  the statement  that the  SLACS lenses  are
representative  of  the  whole  ETGs population  at  these  redshifts.
In both  cases we identify  as SLACS analogues $\sim  18\%$ of
  the initial sample,  proving that the two  dynamical selections give
  similar results.  

\subsection{Properties of the dynamically selected sample}
\begin{figure}
\includegraphics[width=\hsize]{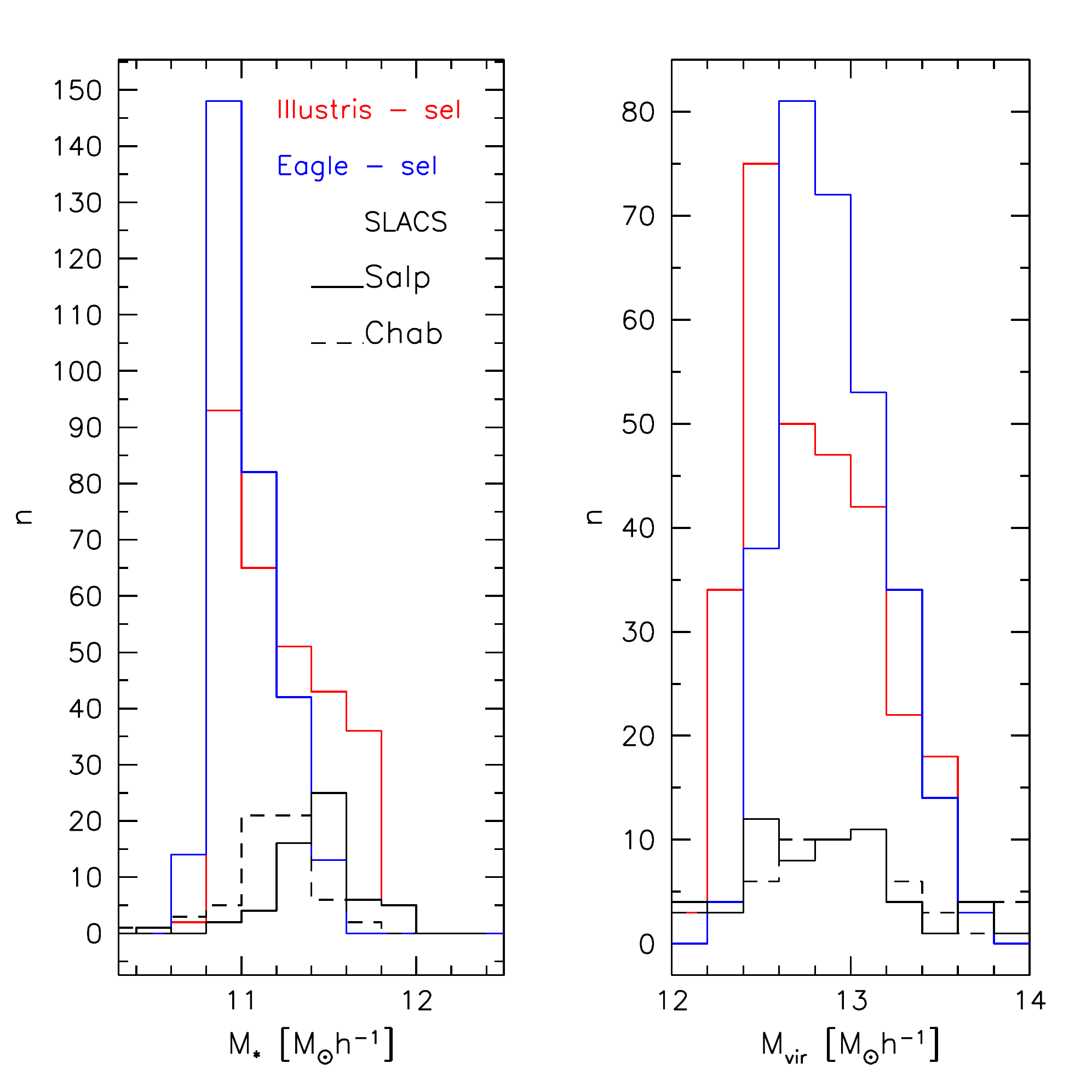}
\includegraphics[width=\hsize]{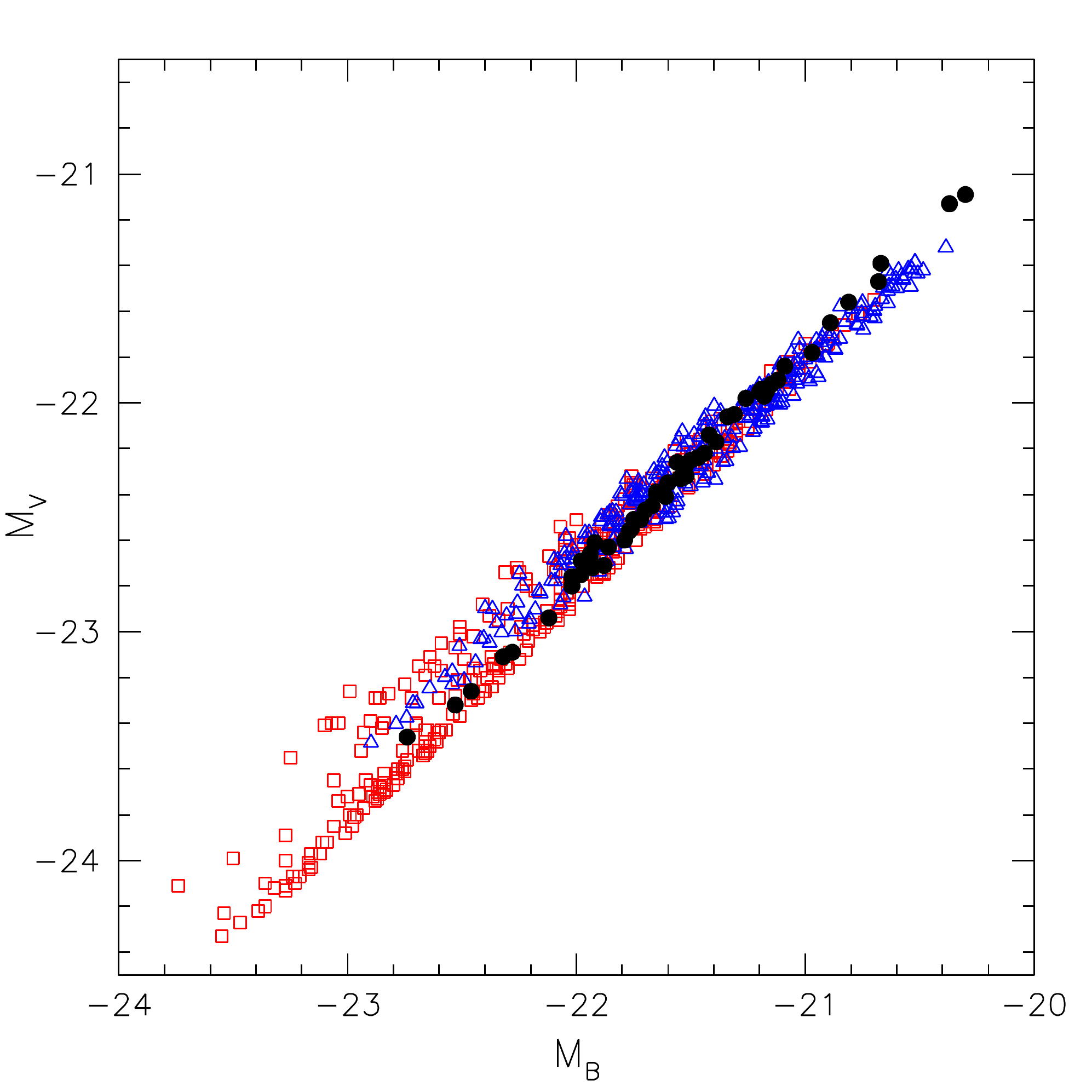}
\caption{$Top$: distribution of stellar mass and virial mass for the SLACS
  galaxies and the samples selected from the simulations; here we plot
  the number of objects found in each case: we have more simulated
  than observed galaxies, but the ranges of stellar and virial masses
  recovered by the selection procedure are in good agreement.
$Bottom$:
  relation between B and V magnitudes for the SLACS lenses and the
  galaxies selected from simulations.
  \label{slacs2}}
\end{figure}

\begin{figure}
\includegraphics[width=\hsize]{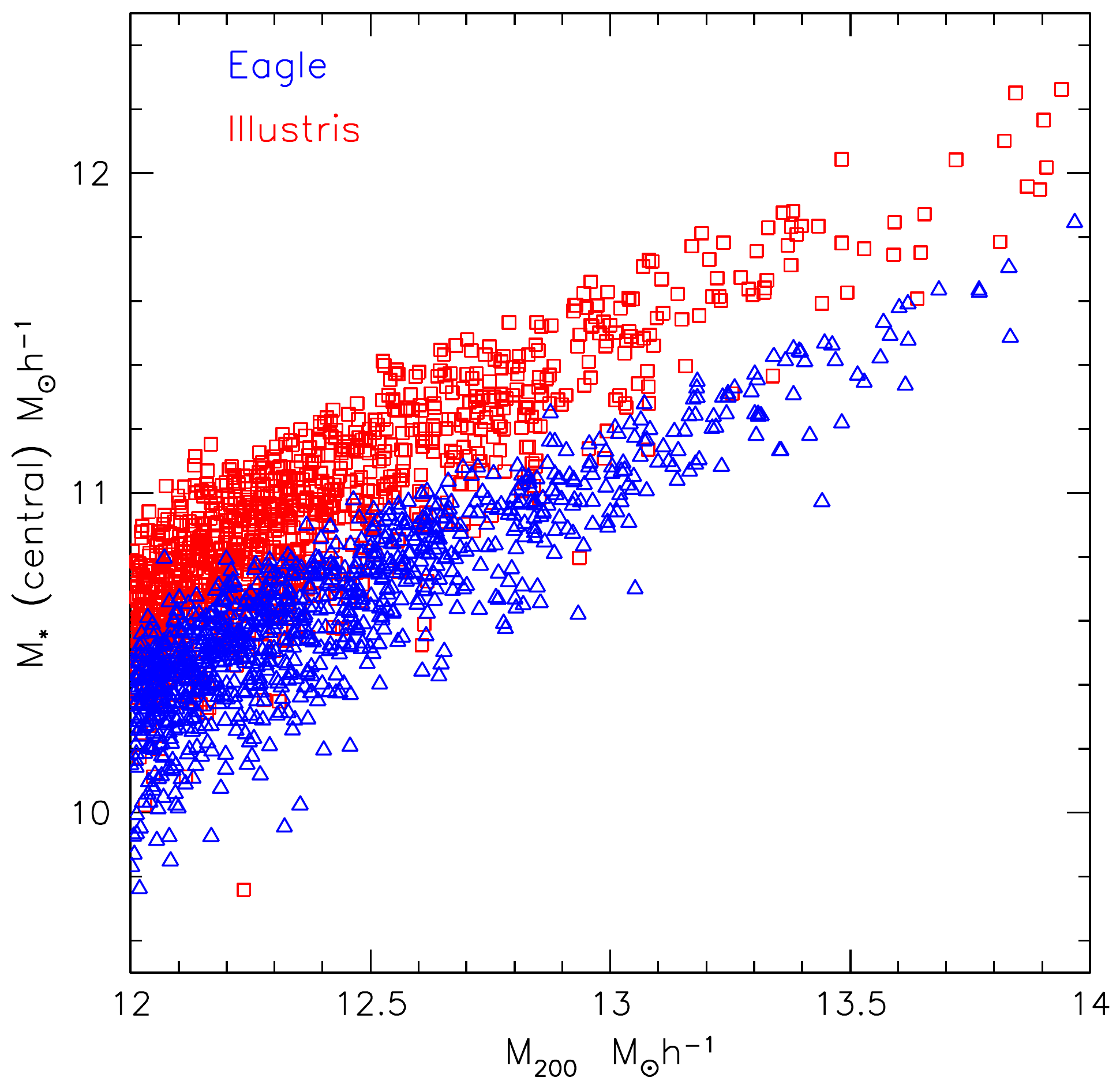}
\caption{Stellar mass of the central galaxy vs. the total mass of
  the halo.\label{mstar}}
\end{figure}

We investigate  other similarities of  the observed galaxies  with the
sample  from  simulations.   \citet{auger10a} provide  a  relation  to
estimate the virial mass, based on a model that relates the virial and
stellar  masses  \citep{moster10},  for various  combinations  of  IMF
(Chabrier, Salpeter  and a Free model)  and halo profile (NFW  and two
profiles from \citealt{blumenthal86} and \citealt{gnedin04} that take into
account  cooling  and  adiabatic contraction).   These  models  define
virial  mass  as the  mass  inside  the  radius enclosing  the  virial
overdensity,  as   calculated  from   the  spherical   collapse  model
\citep{bryan98}.  It is defined in the same way as in the simulations;
we use the virial mass for  the SLACS lenses calculated for a Salpeter
IMF and profile from \citet{gnedin04}.  In Figure \ref{slacs2} we show
the stellar and virial mass distributions of SLACS and of the selected
sample of galaxies,  finding a good agreement.  Here we  note that the
virial mass of the two  simulations peak at slightly different values,
while stellar masses show a better  agreement, bringing us back to the
fact  that Illustris  has a  higher stellar  mass at  fixed halo  mass
(Figure  \ref{mstar}  -  more   details  about  the  different
  composition of central haloes and subhaloes are presented in Appendix
  \ref{Struct}).   Nevertheless, it  is possible  to identify  a good
sample of analogues  in both cases.  In the bottom  panel, we show the
agreement between  the data and  the sample selected  from simulations
for what regards magnitudes, finding a good agreement. As discussed in
\citet{xuD16}, who  selected lens  analogues in a  similar way  in the
Illustris simulation,  simulated early-type galaxies  selected through
velocity dispersion lie well on the fundamental plane.

\section{Comparison with a real detection}\label{sec_det}
In this section we quantify the probability of detecting substructures
in SLACS-like  lenses. Among the  whole SLACS sample, 11  objects have
been  selected  on a  signal-to-noise  basis  and have  been  modelled
searching      for       the      presence       of      substructures
\citep{vegetti14}. These are massive early-type lens galaxies,
  with an average  velocity dispersion of $\sim 270  km/s$ and average
  Einstein radius $<R_{ein}>\sim  4.2 kpc$. In this  sample, only one
lens  has  shown   evidence  of  one  substructure  with   a  mass  of
$3.51\times  10^{9}M_{\odot}$\citep{vegetti10}.   At   the  same  time
non-detections still  carry important information  and can be  used to
constrain the subhalo  mass function. Our aim is to  compare this rate
of  detections with  a prediction  from simulations:  for this
  reason   we  will   calculate   the  probability   of  detecting   a
  substructure,  with  any mass  larger  than  $10^{9}M_{\odot}$ in  a
  random sample  of 11  haloes, chosen  among the  one we  selected as
  SLACS analogues.   We will then  use the projected fraction  of dark
  matter in subhaloes $f_{DM}$ calculated from the analogues, in order
  to compare our estimates with that inferred from observations.

\subsection{Detection probability}

In order to determine the probability of one detection in a sample of
11 lenses, we extract 11 random haloes from our selection and
project their substructures on the image  of each lens.  We repeat the
procedure  600  times  for  each  case (pure  dark  matter,  EAGLE  or
Illustris  hydro),  to  ensure  a  good statistic.   We  find  that  a
substructure - with any mass  larger than $10^{9}M_{\odot}h^{-1}$ - is
present in the sample of 11 lenses  and in the area around the Einstein
rings  where it  could affect  the  lensing signal  in around  41$\%$,
37$\%$ and  $18\%$ of  the cases,  respectively for  dark-matter only,
EAGLE and  Illustris. Among  these, probabilities  are higher  for low
mass  substructures, since  - as  seen in  Figure \ref{ndensity}  - the
scaling between density and mass is of one order of magnitude for each
decade in mass. 

\begin{figure}
\includegraphics[width=0.96\hsize]{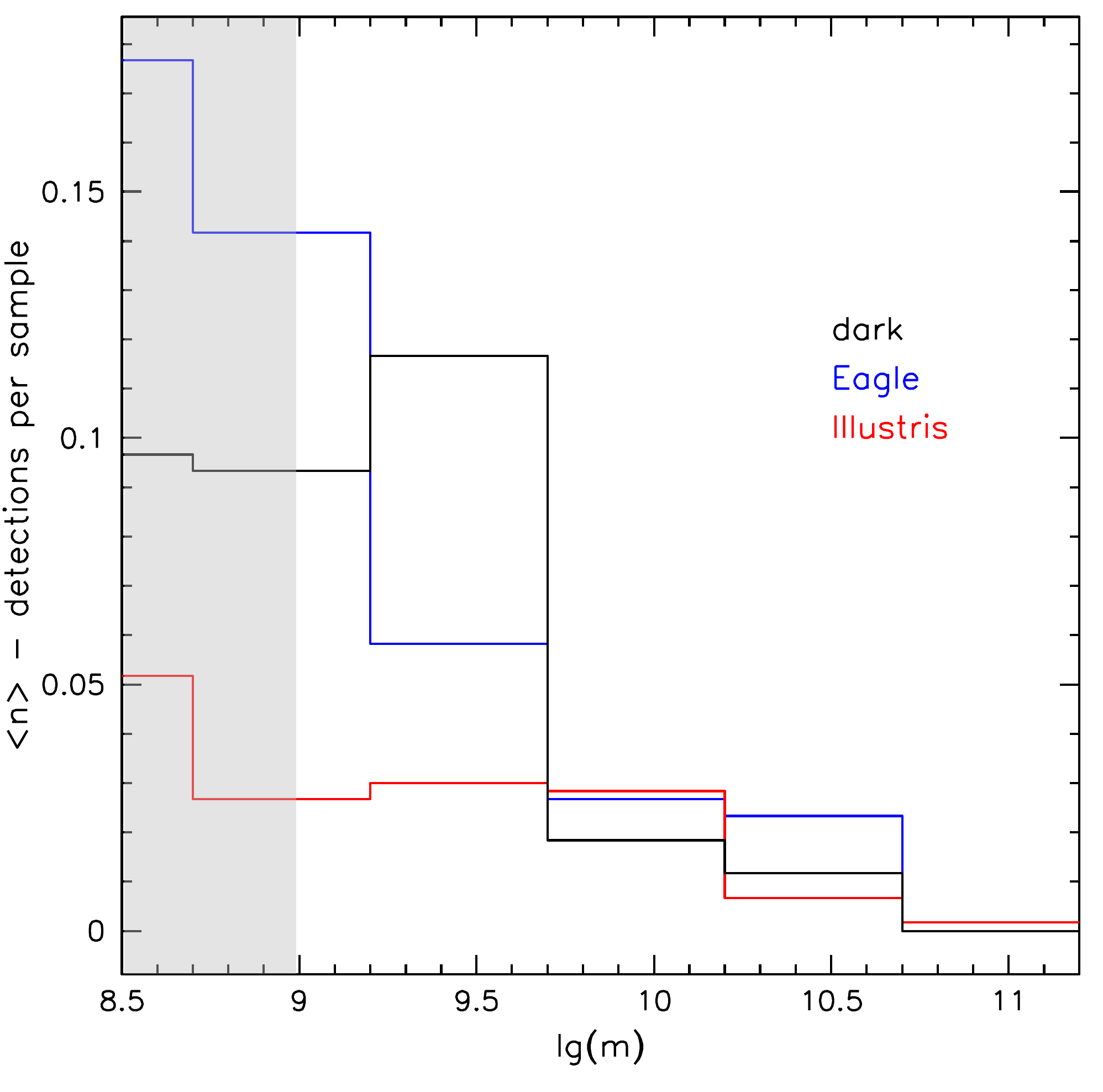}
\caption{Average  number  of  detected  subhaloes  per  sample  of  11
  objects, as  a function of  subhalo mass ($M_{\odot}h^{-1}$). The gray  region indicates
  again where subhaloes have less than 100 particles.\label{sens1}}
\end{figure}

As detailed in \citet{vegetti14},  counting how many substructures lie
in  the  right  region  is  not enough,  because  the  possibility  of
detecting  a   substructure  via  its  gravitational   effect  is  not
independent of its  location: the effect of the  substructure would be
more  evident  for  structured  sources  and in  a  location  where  a
perturbation of the surface brightness could be stronger.  In order to
verify how many substructures can actually be detected, we make use of
the  sensitivity function  maps  created  by \citet{vegetti14}:  these
provide the minimum mass that could  be detected for each pixel in the
imaging of each  of the 11 lenses.  After projecting  and counting the
simulated subhaloes, we associate them  with a pixelized grid with the
same spacing  of the sensitivity function  and we compare the  mass of
the substructure with the sensitivity of the particular pixel it fells
in: if  the lowest  detectable mass  in that pixel  is lower  than the
subhalo mass, then we consider it  a detection, otherwise it is listed
as  non-detection.  This  leaves  us with  an  overall probability  of
detecting one  substructure with mass $M>10^{9}M_{\odot}$  in one lens
and nothing  else in the  others of 20$\%$,  19$\%$ and $10\%$  in the
three cases,  meaning that  nearly $50\%$  of subhaloes  are projected
onto a  pixel that  is not  sensitive enough  to detect  them.  Figure
\ref{sens1} shows the average number of detected subhaloes with a mass
larger than  $10^{9}M_{\odot}h^{-1}$ per  sample of  11 objects,  as a
function  of  subhalo  mass;  probabilities in  wider  mass  bins  are
summarized in Table \ref{tab_sens}.  These results are compatible with
the projected number  counts of Figure \ref{ndensity},  which are then
further reduced by considering the  sensitivity function. We point out
that the sensitivity functions have been calculated using a $10\sigma$
detection as  threshold: using weaker  constraints may lead  to higher
detection probabilities from simulations.   It should be noted
  that these  conclusions are  based exclusively on  projected subhalo
  counts and  we did  not simulate  the actual  lensing effect  of the
  subhaloes, which we plan to explore  in a follow-up paper.  Not only
  the number  of subhaloes is  different in different  simulations, but
  also their structure, profile  and concentration may change, leading
  to a  different gravitational  lensing effect.   Understanding these
  differences  and  the  contribution  of  the  structures  along  the
  line-of-sight may  allow to  go beyond  our results  and effectively
  rule out some hydrodynamical models.

\begin{figure}
\includegraphics[width=\hsize]{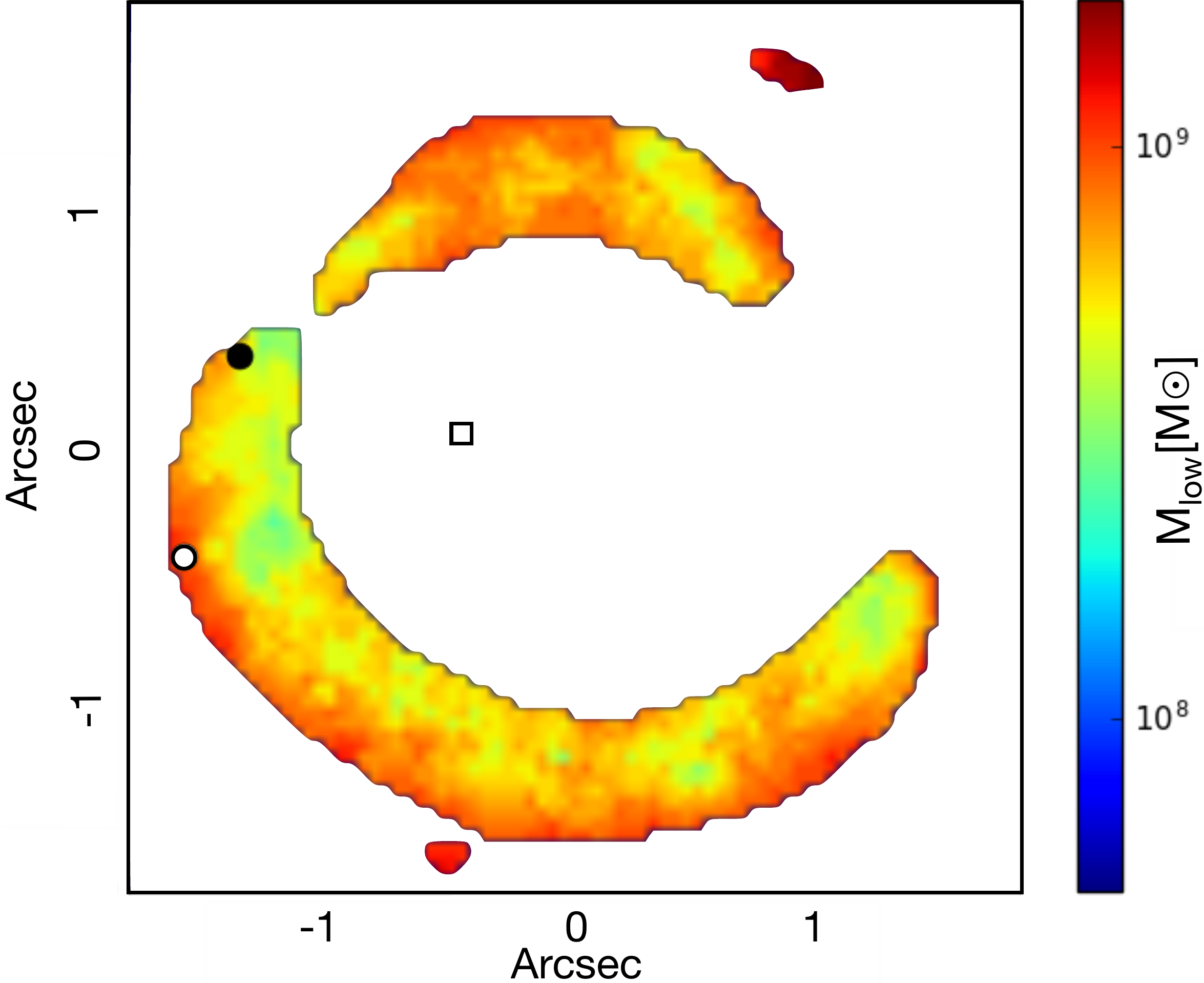}
\caption{Examples of  the projection  of subhaloes on  the sensitivity
  function (lens galaxy SDSSJ0946+1006). The color scale indicates the
  minimum  subhalo mass  that can  be detected  in each  pixel in  the
  region around the Einstein radius  (in $M_{\odot}$). The black point
  shows   and   example   of   detection:   the   detected   mass   is
  $1.6\times 10^{9}M_{\odot}$ - very similar  to the real detection in
  this  region.   The  white  square represent  a  non-detection:  the
  subhalo  mass is  $2.2\times 10^{8}M_{\odot}$,  thus too  low to  be
  detected in  that region.   Finally, the white  circle in  the right
  panel shows a case in which the subhalo falls in the central part of
  the halo but not in the region  around the Einstein radius (it had a
  mass of $9\times 10^{8}M_{\odot}$). \label{sens2}}
\end{figure}

Figure \ref{sens2}  show three examples of  substructures projected on
the  sensitivity function  of  the lens  galaxy SDSSJ0946+1006,  where
there has  been a real  detection \citep{vegetti10}.  The  black point
shows an example of detection: in  the left panel the detected mass is
$1.6\times 10^{9}M_{\odot}$  - very similar  to the real  detection in
this region,  while in the  right panel is  $7\times 10^{8}M_{\odot}$.
The white circle represent a non-detection case, where the mass of the
substructure is lower than the minimum detectable mass, while the white
square point  shows a  substructure that  does not  fall in  the right
region.


\begin{table} \centering
\begin{tabular}{|c|c|c|c|}   \hline 
\multicolumn{4}{c}{ percentage of detections using sensitivity functions}  \\ \hline  
range [$M_{\odot}h^{-1}]$ & DMO &  EAGLE &  Illustris\\  
\hline 
$M>10^{9}$ & 20 & 19 & 10\\
\hline
$10^{9}<M<10^{10}$ & 18.5 & 17 & 8\\
$M>10^{10}$ & 1.5 & 2 & 2\\

 \hline
\end{tabular}
\caption{Probability of detecting one substructure of a certain mass in
  a sample of 11 SLACS-like lenses, taking into account the effect of
  the sensitivity function. Figure \ref{sens1} shows a more detailed
  distribution of the percentage of detections. All these values are
  extracted from 600 random samples from each simulation.\label{tab_sens}}
\end{table}

\subsection{Projected DM fraction $f_{DM}$}
We now use our selection to  calculate the mass fraction in subhaloes:
we calculate the  total projected dark matter mass in  the area of the
sensitivity  functions by  projecting the  dark matter  particles. For
each  of   the  selected  haloes,  we   considered  three  independent
projections for each  of the 11 sensitivity  functions.  The resulting
projected  dark  matter masses  are  generally  consistent with  those
calculated for  this sample of 11  SLACS lenses and on  average of the
order of  $2.5 \times  10^{10}M_{\odot}h^{-1}$. We then  calculate the
expected number  density of  subhaloes $dn/(kpc/h)^{2}$, as  in Figure
\ref{ndensity}  for   subhaloes  with  masses  between   $10^{8}$  and
$5\times  10^{9}M_{\odot}h^{-1}$; since  this scales  approximately of
one order  of magnitude for  each decade  in mass, we  extrapolate the
expected  number   of  lower  mass  subhaloes   between  $10^{6}$  and
$10^{8}M_{\odot}h^{-1}$ that we do not  resolve in the simulations. We
then use the  total projected dark matter mass and  the number density
of subhaloes  in the sensitivity function  areas, obtained multiplying
the mean value for the total area. The mean densities in substructures
in the  range $4\times  10^{6}-4\times 10^{9}M_{\odot}h^{-1}$  and the
resulting  mean mass  fraction in  substructures on  the areas  of the
sensitivity functions  are listed  in Table \ref{tab_sens2}.   For the
dark  matter only  case we  are consistent  with a  value of  $f_{DM}$
compatible  with  \citet{vegetti14} -  and  thus  consistent with  the
predictions from \citet{xuD15}.  The  mass fraction in substructure in
the  range  $10^{6}-10^{9}M_{\odot}h^{-1}$  is   then  lower  for  the
simulations with baryons and  especially for the Illustris simulation.
These values,  combined with the  best fit  slope of the  subhalo mass
function  $\alpha$ can  be used  to fully  constrain the  subhalo mass
function and  can be compared  with those inferred  from observations.
\citet{vegetti14} found a mean substructure projected mass fraction of
$f_{DM}=0.0076^{+0.0208}_{-0.0052}$  for a  uniform prior  on $\alpha$
and $f_{DM}=0.0064^{+0.0080}_{-0.0042}$ for a gaussian prior with mean
1.9   and    standard   deviation   0.1.    Thus,    the   values   of
($f_{DM}$,$\alpha$) from  the DMO  and from the  EAGLE hydro  runs are
compatible with  their findings within  the errors, both for  the mass
fraction  $f_{DM}$  and the  slope  $\alpha$.   The results  from  the
Illustris hydro run  instead do not lie within  the errors. If
  we use the  double power-law fit to the subhalo  mass function (from
  Table \ref{tab_sim1}), we obtain  slightly higher values of $f_{DM}$
  for  the  hydro  runs;  even   in  this  case,  the  combination  of
  ($f_{DM}$,$\alpha$)  obtained  from  EAGLE is  compatible  with  the
  observational results  within the  errors, while that  obtained from
  Illustris is not,  leaving our conclusions unchanged.  We prefer the
  results  obtained with  a single  power-law mass  function for  this
  comparison, in  order to be  consistent with  the model used  in the
  lensing analysis.    We compare  with this prediction and  not with
those  of \citet{vegetti10}  and  \citet{vegetti12},  since the  first
included only one lens and the second  is based on a lens which is not
part  of the  SLACS sample  and  has therefore  a different  selection
criteria and a different mass and redshift.

\begin{table} \centering
\begin{tabular}{|c|c|c|}   \hline 
\multicolumn{3}{c}{ subhalo mass fraction}  \\ \hline  
sim & $<\rho_{sub}>[M_{\odot}h/kpc^{2}]$&  $f_{DM}$\\  
\hline 
DMO & $5.826\times 10^{6}$ & 0.0044$\pm$ 0.0018\\
EAGLE & $3.858\times 10^{6}$ & 0.0025$\pm$0.0012\\
Illustris & $1.541\times 10^{6}$ & 0.0012$\pm$0.0004\\

 \hline
\end{tabular}
\caption{Average projected mass density in subhaloes on the areas of
  the sensitivity function, for the simulated analogues of SLACS lenses. \label{tab_sens2}}
\end{table}

\section{Summary} \label{conclusion}

We have analysed the results of the two most recent simulations (EAGLE
and  Illustris), aiming  to  characterize the  subhalo population  in
simulations with different baryonic  physics models. We concentrate on
haloes  mass  between  $10^{12.5}$  and  $10^{14}M_{\odot}h^{-1}$  and
redshift  between  0.2  and  0.5,   since  we  want  to  compare  with
observations of ETGs  at these redshifts.  Here we  summarise our main
results:
\begin{itemize}
\item the presence of baryons modifies the abundance and structure of
  haloes, through  processes such as adiabatic  contractions, cooling,
  stellar and AGN  feedback. As a consequence,  the subhalo population
  is affected $(i)$ by the different  abundance of haloes in the field
  that can  be accreted  by larger  haloes and  $(ii)$ by  a different
  dynamic and survival of substructure inside the main halo. Depending
  on the adopted physical model, the  depletion in the low-mass end of
  the subhalo mass  function changes: in the EAGLE hydro  run, we find
  $\simeq  20\%$ fewer  subhaloes  with a  mass  between $10^{8}$  and
  $10^{10}M_{\odot}$,   while  for   the  Illustris   simulation  this
  percentage can be  as high as $40\%$. A different  effect is present
  also at  higher subhalo masses  ($10^{11}-10^{12}M_{\odot}$), where
  Illustris shows  an excess of subhaloes  in the hydro run,  which is
  not   present  in   EAGLE  (Figure \ref{massfunc});   this  kind   of
  differences need to  be investigated more since they  may be similar
  to those caused by warm  dark matter models \citep{lovell14,li16} at
  the low mass end.
\item We model the subhalo mass function for different halo masses,
  using a relation from \citet{giocoli08a}; for the DMO case, we find
  a slope $\alpha=-0.9$, consistent with previous studies, while we
  find shallower slopes for the hydro runs, with  $\alpha=-0.85$ for
  the EAGLE hydro run and  $\alpha=-0.76$ for the Illustris one
  (Figure \ref{massfunc2});
\item the  projected number density  of subhaloes  is quite flat  as a
  function  of  radius,  as  shown  already  by  \citet{xuD15}  for  a
  different mass range;  the abundance of subhaloes that  can be found
  in projection in the central regions  of the halo decreases by about
  one order of magnitude for each decade in subhalo mass.

\end{itemize}

We conclude that baryonic physics has  an important impact on the halo
structure and on  the subhalo population; this needs to  be taken into
account   when   we   compare  predictions   from   simulations   to
observational results.  The reduction in the number of small subhaloes
is a clear  consequence of the presence of baryons  and of stellar and
AGN  feedback. However,  in  order to  distinguish  this effect  from
others, such as that of warm dark matter models \citep{lovell14,li16},
we need  to reach  lower subhalo  masses. We  plan to  investigate these
differences  with zoom-in  high-resolution simulations  in a  follow-up
paper.

The  second  part  of  this   work  focuses  on  the  comparison  with
observational results,  and in  particular with  the SLACS  lenses. We
searched  for analogues  in  the simulations,  considering ETGs  which
match the properties of the SLACS  galaxies. We found a good number of
these  analogues  at  redshift  between 0.2  and  0.5,  selecting  the
galaxies  by dynamical  properties  and velocity  dispersion; we  then
verify  that the  selected galaxies  lie in  the right  region of  the
$M_{*} - r_{e}$ plane and that  the distribution of total stellar mass
and virial  mass are  consistent with  the observed  ones. We  use the
selected  galaxies to  estimate subhalo  detection probabilities  with
different physical models.  For this,  we extract random samples of 11
SLACS-like haloes from the  simulations and project their substructure
on the sensitivity functions of the  real lenses, in order to estimate
how many  subhaloes could be  detected.  We conclude that:

\begin{itemize}
\item  1 detection
with a  mass $M>10^{9}M_{\odot}$  in a  sample of 11  lenses is  not a
certain event,  and it  has a  probability of 20  $\%$ (DMO),  19 $\%$
(EAGLE) or 10 $\%$ (Illustris);
\item many  more observed lenses  of ETG mass  are needed to  ensure a
  good number of detections and thus being able to fully constrain the
  subhalo mass  function;
\item the  dark matter fraction in  subhaloes within the areas  of the
  sensitivity functions,  for the three  models is $f_{DM}$(DMO,
  EAGLE,   Illustris)  =   (0.0044,0.0025,0.0012).    The  values   of
  ($f_{DM}$,$\alpha$) from the  DMO and from the EAGLE  hydro runs are
  both compatible  with the  findings of \citet{vegetti14}  within the
  errors,  while those  from the  Illustris are  significantly
    lower
\end{itemize}

This  clearly  shows that  substructure  lensing  not only  allows  to
distinguish between  different dark  matter models, but  also between
feedback and  galaxy formation models, provided  that the contribution
from the line-of-sight structures  is well understood. This last aspect will
be the main focus of a follow-up paper. Another scenario that needs to
be addressed  is that of different  warm dark matter models,  since it
also causes  a lack of low  mass substructures; we plan  to extend our
findings to this  and and to the combination between  warm dark matter
and baryons in  a future work, using higher  resolution simulations in
order to reach lower masses.

\section{Acknowledgments}

We thank  Dandan Xu, Simon  White, Joop Schaye, Carlo  Giocoli, Carlos
Frenk, Rob Crain  and Tom Theuns for useful  discussions and comments.
We thank Mark Lovell for  providing some additional galaxy catalogues.
We thank  the EAGLE collaboration for  granting us access to  the data
for  this project.   For part  of this  work, we  used the  DiRAC Data
Centric system  at Durham  University, operated  by the  Institute for
Computational  Cosmology on  behalf  of the  STFC  DiRAC HPC  Facility
(www.dirac.ac.uk).   This   equipment  was  funded  by   BIS  National
E-infrastructure  capital  grant  ST/K00042X/1,  STFC  capital  grants
ST/H008519/1   and   ST/K00087X/1,   STFC   DiRAC   Operations   grant
ST/K003267/1  and Durham  University. DiRAC  is part  of the  National
E-Infrastructure. We thank the  anonymous referee for his/her valuable
comments that have helped us to improve the quality of the paper.

\appendix \newpage

\section{Structure of subhaloes}\label{Struct}

\begin{figure*}
\includegraphics[width=0.65\hsize]{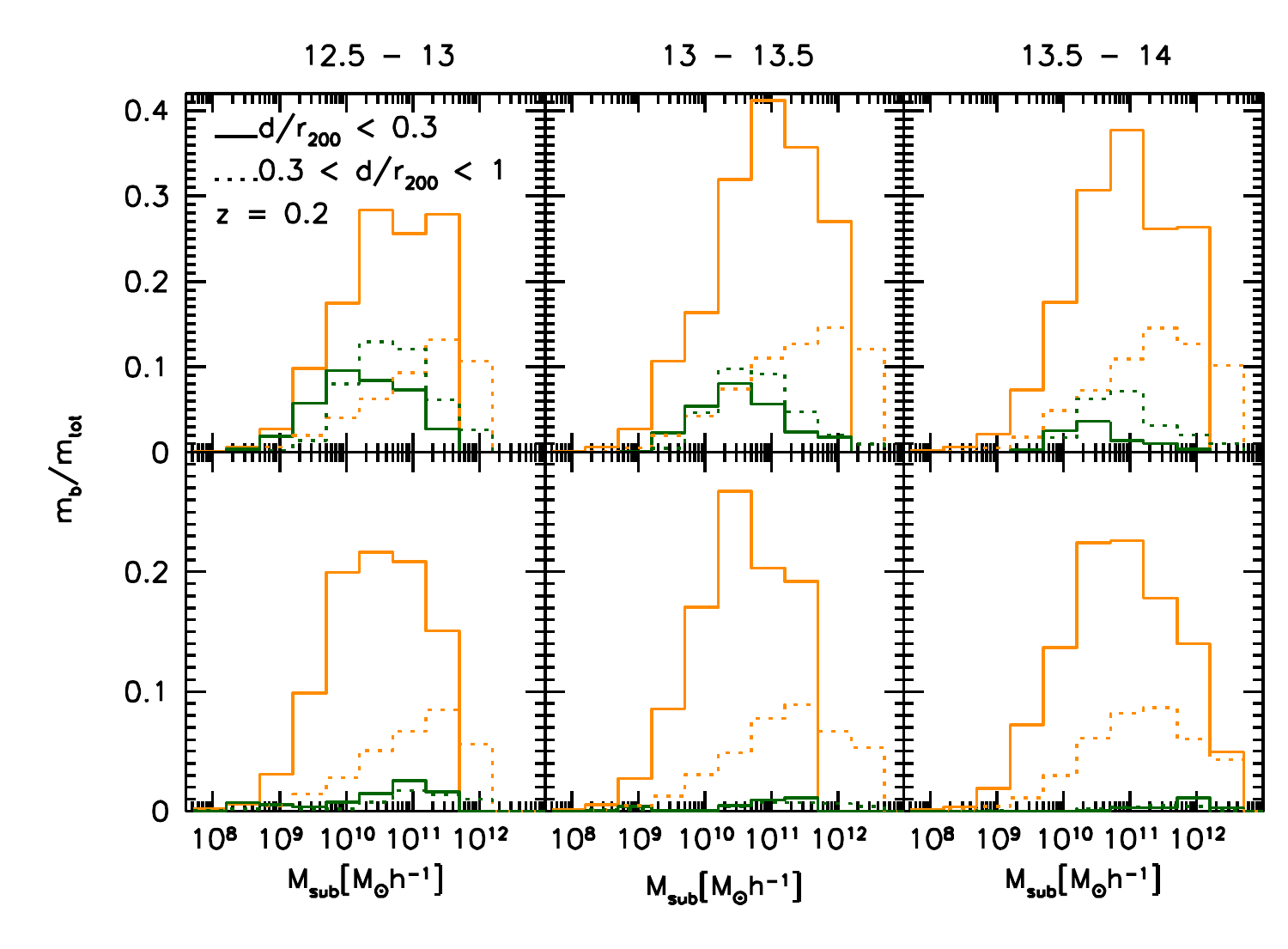}
\includegraphics[width=0.29\hsize]{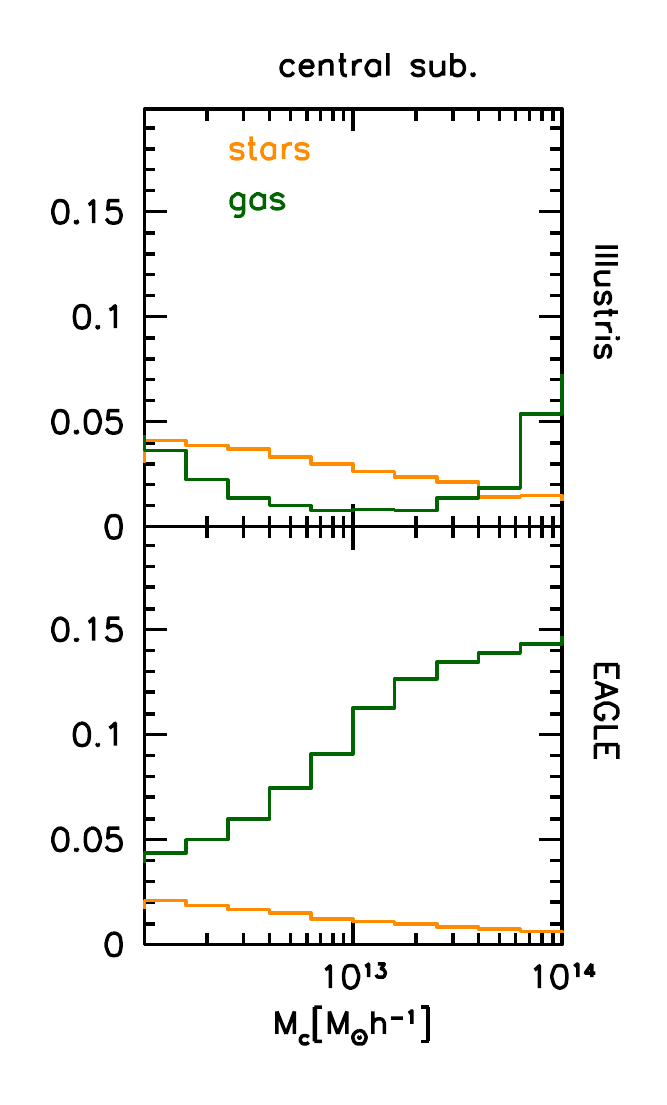}
\includegraphics[width=0.65\hsize]{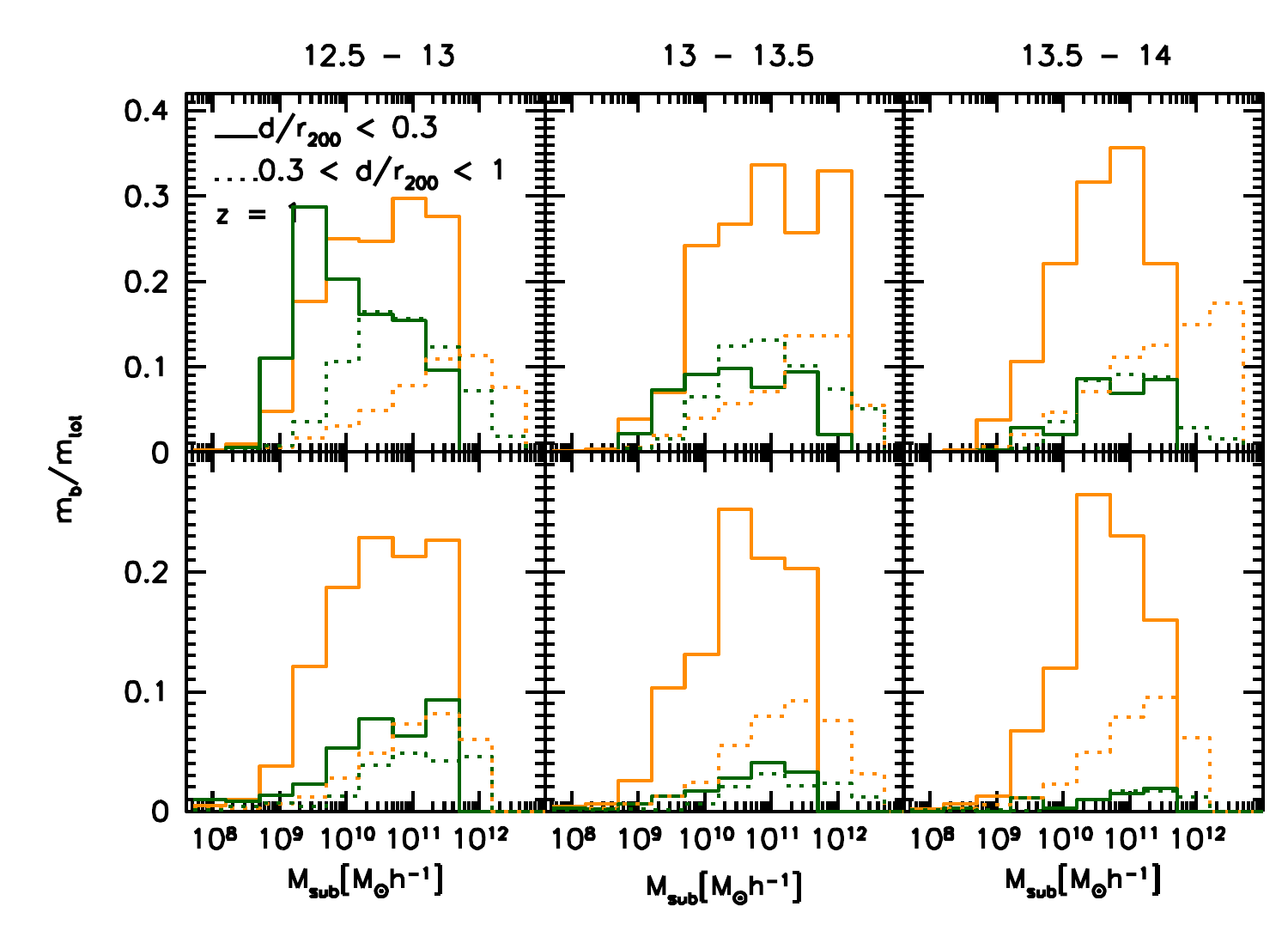}
\includegraphics[width=0.29\hsize]{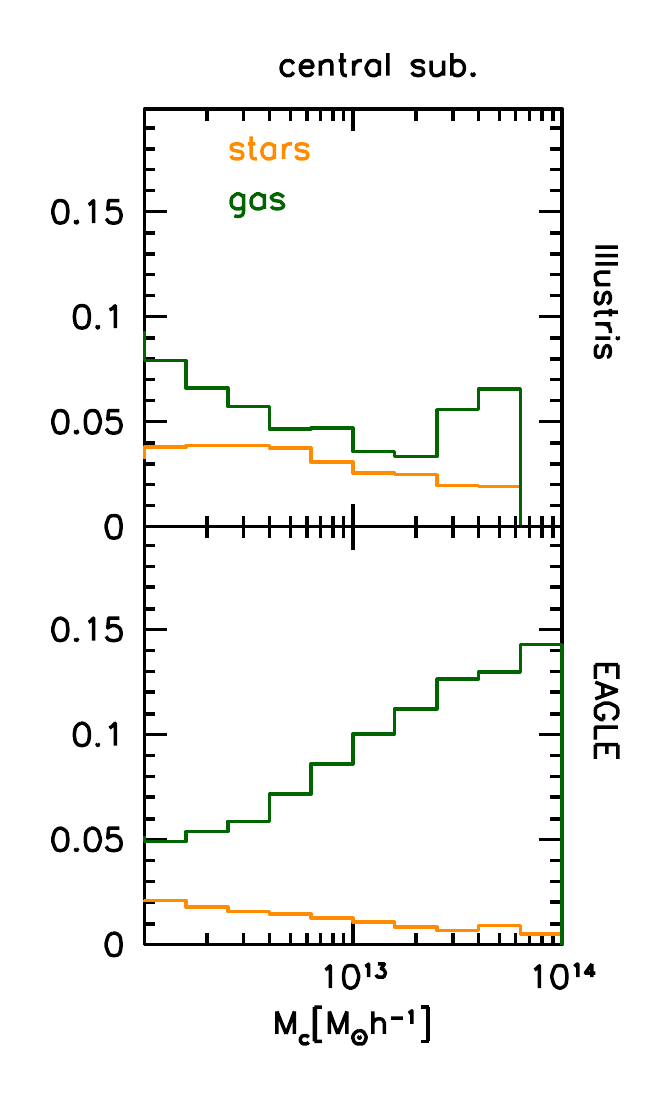}
\caption{ $Left$:  subhalo composition. In  all panels the  orange and
  green lines show, respectively, the mass fraction in star and gas of
  the subhaloes.  Subhaloes are  binned according to  the mass  of the
  parent halo; each column shows the results for one of the three bins
  ($12.5\leq  lg(M_{200})\leq 13$,  $13\leq lg(M_{200})\leq  13.5$ and
  $13.5\leq lg(M_{200})\leq 14$). Subhaloes are also divided according
  to their  distance from the  centre of  that halo: solid  lines show
  those which  lie within 30\%  of the radius $r_{200}$,  while dotted
  lines  those   which  are   find  within  $0.3\times   r_{200}$  and
  $r_{200}$.  Even  though  low   mass  subhaloes  are  mainly
    completely dark,  we want to  stress that the baryonic  content is
    not      well     resolved      for     masses      lower     than
    $10^{9}M_{\odot}h^{-1}$. $Right$: composition of the central halo
  (first subhalo or smooth component of the halo).\label{composition}}
\end{figure*}

We look  at the  structure of  subhaloes and host  haloes in  the full
hydrodynamical runs  for the  lens analogues.  SUBFIND  identifies the
main  ``smooth'' component  of  the FOF  halo as  the  first and  most
massive  subhalo,  followed  in  the  catalogue  by  all  the  smaller
structures.  In Figure \ref{composition}  we distinguish the main halo
and  its subhaloes  and  study  their baryonic  content:  we plot  the
average percentage of mass built up  by stars and gas, respectively in
orange  and green.   The  right  panel shows  the  composition of  the
central  subhalo (i.e.   the main  halo) for  the two  simulations, at
redshifts 0.2  and 1.   The results from  the two  simulations present
significant differences: in the EAGLE run, the main halo contains much
more gas than in the Illustris and  at the same time less gas is bound
to the subhaloes.   Moreover, we note again a higher  stellar mass in
Illustris   galaxies,  as   in  Figure   \ref{mstar}.   As   shown  in
\citet{schaye15},  the  star  formation  is higher  in  the  Illustris
simulations for all  masses and the feedback model  induces a stronger
AGN feedback, which  expels almost all the gas from  the halo with the
purpose of quenching star formation - and is one of the known problems
of the Illustris  recipe.  The effect of different  feedback models on
the baryon and gas fractions  has been studied in \citet{velliscig14},
who  found an  important  depletion  in the  presence  of  an AGN.   A
selection of  the simulated  galaxies using  observational constraints
(as stellar mass, effective radius  or magnitude) may thus be affected
by the different composition of the central halo.

The left panels show the same  for the subhaloes, binned by 
halo  mass (different  columns) and  distance from  the centre:  the
subhaloes  which lie  in the  very centre  of the  halo -  closer than
$0.3\times r_{200}$ - are represented by solid lines, while the others
- between $0.3\times  r_{200}$ and $r_{200}$ - by  dotted lines. First
of all, we find  - as in previous works -  that the smallest subhaloes
($10^{8} -  10^{9}$ $M_{\odot}h^{-1}$) are almost  completely dark and
do not form stars.  Moreover, subhaloes which lie near the centre lost
the majority of  their gas, so that  stars build up to 40  $\%$ of the
total  mass,  while  more  distant  satellites  show  fewer  signs  of
stripping.

\begin{figure}
\includegraphics[width=\hsize]{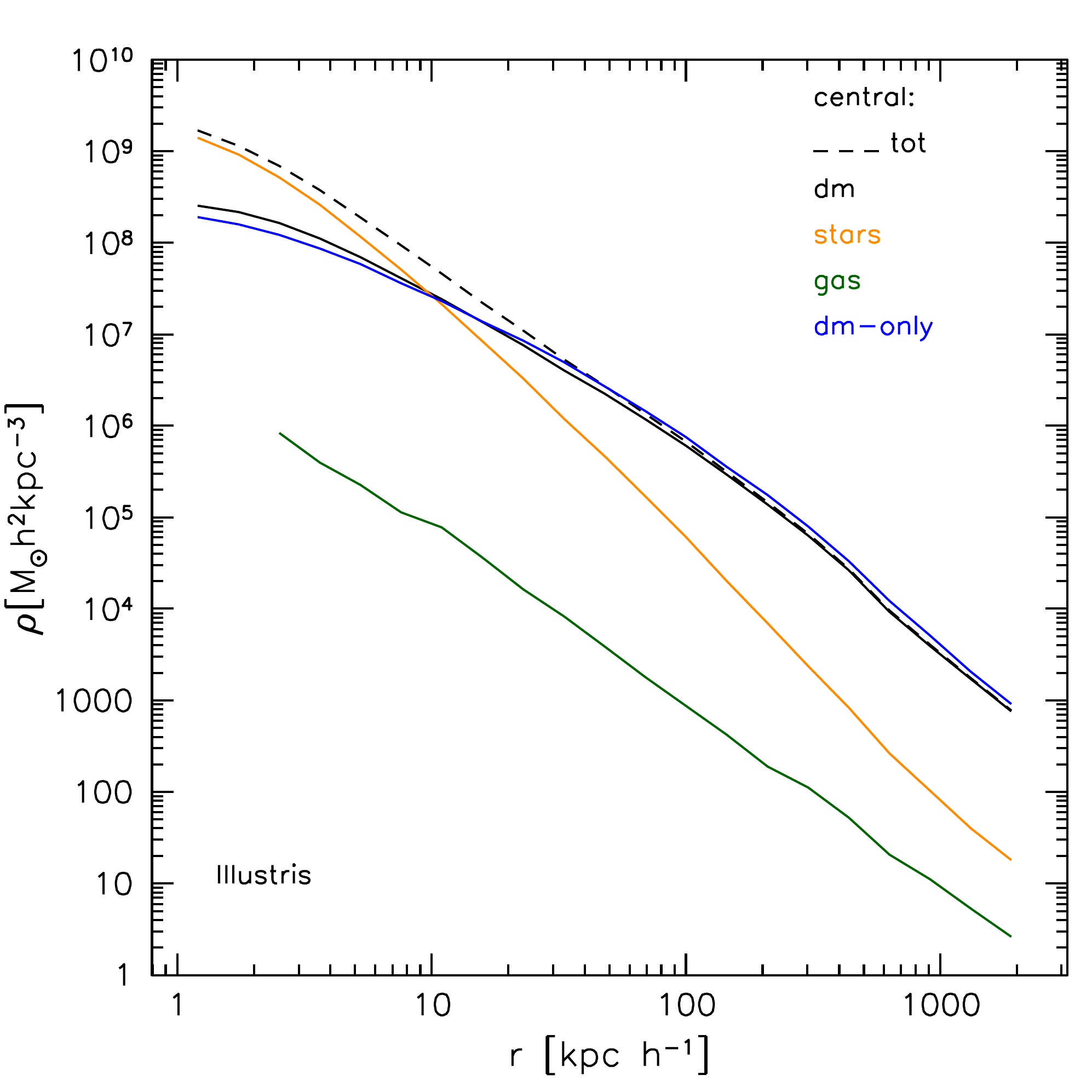}
\includegraphics[width=\hsize]{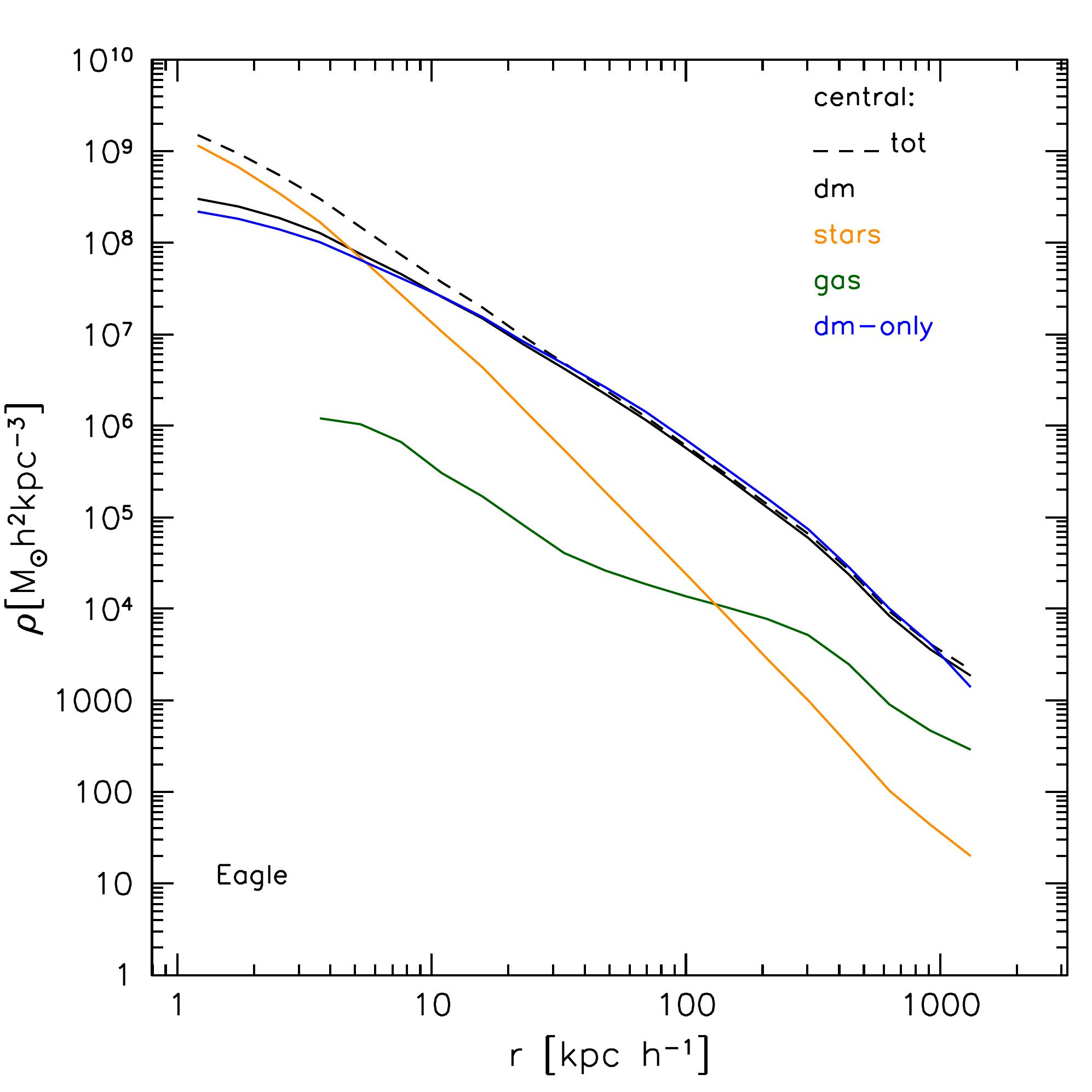}
\caption{  Mean   density  profile   of  the  central   halo  (without
  subhaloes). We chose 10 haloes from each of the two hydro runs: they
  have  $M_{200}\simeq   10^{13}M_{\odot}h^{-1}$  and   very  similar
  dynamical properties;  the central galaxy  has been identified  as a
  massive  elliptical. We show the profile of each component (dark
  matter - black,  stars - orange, gas - green)  and the total density
  profile in black  dashed lines. The blue curve shows  the profile of
  the counterparts  of these haloes in  the dark matter only  run.  As
  from Figure  \ref{composition} the central halo  from EAGLE contains
  more  gas, while  the  one  from Illustris  has  a  bit more  stars;
  nevertheless,  the  total and  the  dark  matter profiles  are  very
  similar between the two simulations. \label{profile_c}}
\end{figure}

Figure \ref{profile_c}   shows the mean  radial density
profiles of  the central  haloes.  The mean density profile of  the central is very similar in the
two cases (Figure \ref{profile_c}): stars and gas behave differently,
but  their contributions  sum up  to give  a comparable  total density
profile.  

We  plan  to analyse  subhalo  profiles  and concentration  in
  detail  in  a follow-up  paper,  using  ray-tracing to  model  their
  influence on  the lensing  signal. Differences  due to  the baryonic
  physics implementation  may arise and analysing  possible systematic
  differences is important,  as subhalo concentration plays  a role in
  the possibility to observe them through gravitational lensing.

\bibliographystyle{mn2e}
\bibliography{paper.bbl}
\end{document}